\documentclass[preprint,aps,12pt,nofootinbib,superscriptaddress]{revtex4}
\pdfoutput=1

\usepackage[pdftex]{color,graphicx}
\usepackage{mathrsfs}
\usepackage{amssymb}
\usepackage{amsmath}
\usepackage{mathtools}
\usepackage{epsfig}
\usepackage{graphics}
\usepackage{bbm}
\usepackage{slashed}

\newcommand{\be}{\begin{equation}}
\newcommand{\ee}{\end{equation}}
\newcommand{\bea}{\begin{eqnarray}}
\newcommand{\eea}{\end{eqnarray}}

\newcommand{\ba}{\begin{array}}
\newcommand{\ea}{\end{array}}

\begin{document}

\vspace*{-15mm}
\begin{flushright}
SISSA 28/2015/FISI\\
CP3-Origins-2015-024 DNRF90\\
DIAS-2015-24\\
IPMU15-0092
\end{flushright}
\vspace*{0.7cm}

\title{Radiative Corrections to Light Neutrino Masses\\ 
in Low Scale Type I Seesaw Scenarios \\
and Neutrinoless Double Beta Decay}

\author{J. Lopez-Pavon}
\email[]{jlpavon@sissa.it}
\affiliation{SISSA and INFN - sezione di Trieste, 
via Bonomea 265, 34136 Trieste, Italy.}
\author{E. Molinaro}
\email[]{molinaro@cp3.dias.sdu.dk}
\affiliation{CP$^3$-Origins and Danish Institute for Advanced Study, 
University of Southern Denmark,
Campusvej 55, DK-5230 Odense M, Denmark}
\author{S. T. Petcov
\footnote{Also at:
 Institute of Nuclear Research and Nuclear Energy,
  Bulgarian Academy of Sciences, 1784 Sofia, Bulgaria.} 
}
\affiliation{SISSA and INFN - sezione di Trieste, 
via Bonomea 265, 34136 Trieste, Italy.}
\affiliation{Kavli IPMU (WPI), University of Tokyo, 
5-1-5 Kashiwanoha, 277-8583 Kashiwa, Japan}

\begin{abstract}

We perform a detailed analysis of the one-loop corrections to the light neutrino mass matrix within low scale type I seesaw extensions of the
Standard Model and their implications in experimental searches for neutrinoless double beta decay. We show that 
a sizable contribution to the effective Majorana neutrino mass from the exchange of heavy Majorana neutrinos is always possible,
provided one requires a fine-tuned cancellation between the tree-level and one-loop contribution to the light neutrino masses.
We quantify the level of fine-tuning as a function of the seesaw parameters and introduce a generalisation of the Casas-Ibarra parametrization of the neutrino Yukawa matrix,
which easily allows to include the one-loop corrections to the light neutrino masses.

\end{abstract}

\maketitle

\section{Introduction}

The main goal of this work is to study in detail 
under which conditions the right-handed (RH) neutrinos present
in a general type I seesaw scenario \cite{seesaw} can give a direct sizable 
contribution to the neutrinoless double beta ($0\nu\beta\beta$) 
decay rate, i.e., a contribution in the range of sensitivity of the current 
and upcoming $0\nu\beta\beta$ decay experiments,
once all the relevant constraints are included in the analysis.

In~\cite{Blennow:2010th,LopezPavon:2012zg}, it was shown that 
a sizable sterile neutrino
contribution to the $0\nu\beta\beta$ decay can be achieved 
if the heavy neutrino spectrum is 
hierarchical, with at least one RH neutrino with mass 
$M$ below $100$ MeV and the other state(s) above 
this scale. However, this spectrum is disfavoured 
by cosmological observations since the region 
$M\in[1$ eV, $100$ MeV$]$ is excluded
by BBN and CMB data~\cite{Hernandez:2013lza,Hernandez:2014fha}. 
In~\cite{Ibarra:2010xw,Ibarra:2011xn,Mitra:2011qr} 
the possibility of having a relevant contribution 
from heavy RH neutrinos up to the TeV
scale was explored.~\footnote{The interplay between the light and
heavy Majorana neutrino contributions in $0\nu\beta\beta$ decay was investigated phenomenologically
first in \cite{HPR83}.}~It was found that indeed RH neutrinos as heavy as 
$100$ GeV$\textendash$10 TeV could, in principle, give a 
sizable and observable contribution to the $0\nu\beta\beta$ decay rate. 
In~\cite{Mitra:2011qr} the role of the fine-tuning and one-loop effects were discussed, 
concluding that for RH neutrino masses above $10$ GeV a relatively high level of fine-tuning 
would be required. In~\cite{LopezPavon:2012zg} a more detailed study of 
the one-loop effects was performed and it was found that indeed 
they are significant and can play a very important role in the type I seesaw scenario. 
The lepton number violation introduced through the RH neutrino Majorana mass term, 
required to obtain a sizable effect in the $0\nu\beta\beta$ decay rate, naturally 
appears at one-loop level in the light neutrino sector. 
If fine-tuning is not invoked, the light neutrino mass constraints on 
the one-loop corrections make it very difficult to obtain a 
significant (RH) heavy Majorana neutrino contribution  
in the $0\nu\beta\beta$ decay
 effective Majorana mass, i.e., to have $|m_{\beta\beta}^{\text{heavy}}|\gtrsim 0.01$ eV,
$m_{\beta\beta}^{\text{heavy}}$ being the heavy 
Majorana neutrino contribution under discussion.
We will show, in particular, that the scenario in which RH neutrinos 
with a mass $M\gtrsim 1$ GeV can give a sizable contribution
to the $0\nu\beta\beta$ decay rate necessarily involves a fine-tuned 
cancellation between the tree-level and one-loop light neutrino contributions.

More specifically, in this work we re-analyse the conditions under which 
the heavy Majorana neutrinos with masses $M >100$ MeV 
of the type I seesaw scenario 
can give a significant direct contribution to
the  $0\nu\beta\beta$ decay effective Majorana mass, 
i.e., a contribution in the range of sensitivity of the current 
and upcoming $0\nu\beta\beta$ decay experiments.   
We  show that for $M \gtrsim$ a few GeV  
this requires a relatively large active-sterile neutrino mixing 
(charged current couplings of the heavy Majorana neutrinos). 
We clarify which seesaw realisations can provide 
the requisite mixing. We discuss the impact of the 
one-loop corrections in the different type I seesaw 
realisations considered. We analyse also numerically 
the problem of the sizable heavy Majorana neutrino contribution 
to the  $0\nu\beta\beta$ decay effective Majorana
mass,  by studying the full parameter space, including the 
relevant one-loop corrections and the bounds on the 
active-sterile neutrino mixing from direct searches, charged lepton flavour violation 
and non-unitarity~\cite{Antusch:2006vwa,Antusch:2008tz,Atre:2009rg,Alonso:2012ji,Antusch:2014woa,Drewes:2015iva,Dinh:2012bp,Cely:2012bz}. We 
quantify, in particular, the level of fine-tuning required in order to have a 
sizable heavy neutrino contribution to the $0\nu\beta\beta$ 
decay rate. In order to do the analysis and 
generate the right pattern 
for the light neutrino masses and mixing, 
we have constructed a modification of the Casas-Ibarra 
parametrization~\cite{Casas:2001sr}, 
which takes into account the impact of the
one-loop corrections.

The paper is organized as follows: in section~\ref{sec1} we derive under which conditions it is possible to obtain a sizable active-sterile neutrino mixing,
which can strongly affect the effective Majorana neutrino mass, $m_{\beta\beta}$. In section~\ref{sec3} we study the impact on $m_{\beta\beta}$ of the one-loop 
corrections to the light neutrino masses and present our modified Casas-Ibarra parametrization which takes into account the one-loop effects. In 
section~\ref{BetaBeta} we perform the numerical analysis and quantify the level of fine-tuning necessary to have a dominant contribution in $m_{\beta\beta}$ from the exchange of the heavy (sterile) neutrinos. Finally, we summarise our results in the concluding section.

\mathversion{bold}
\section{Large Active-Sterile Neutrino 
Mixing and $0\nu\beta\beta$ Decay}
\mathversion{normal}
\label{sec1}

We consider the most general type I seesaw scenario \cite{seesaw} 
with $n\geq2$ RH neutrino fields $\nu_{sR}$ ($s=1,\ldots,n$). 
After the spontaneous breaking of the electroweak (EW) symmetry the 
full neutrino mass Lagrangian is
\begin{equation}
\mathcal{L}_{\nu}\;=\; -\, \overline{\nu_{\ell L}}\,(m_{D})_{\ell s}\, \nu_{sR} - 
\frac1 2\, \overline{\nu^{c}_{sL}}\,(M_{R})_{st}\,\nu_{tR}\;+\;{\rm h.c.}
\label{typeI}
\end{equation}
%
where $\ell=e,\mu,\tau$ and $\nu^{c}_{sL}\equiv C\, \overline{\nu_{sR}}^T$, 
$C$ being the charge 
conjugation matrix. $M_{R} = (M_{R})^T$ is the Majorana mass matrix of the 
RH neutrinos and $m_{D}$ is the $3\times n$ neutrino Dirac mass matrix.
The full mass matrix derived  from Lagrangian (\ref{typeI}) is therefore
\be
 \mathcal{M} \equiv \begin{pmatrix} \mathbf{O} &  m_D \\
 m_D^T & M_R  \end{pmatrix} = U^* \,\text{diag}\left(m_i,M_k\right)U^\dagger,
\label{Mnu}
\ee
%
where $m_i$ ($i=1,2,3$) and $M_k$ ($k=1,\ldots,n$) are the 
light and heavy Majorana neutrino masses, respectively. 
We define $\mathbf{O}$  as a $3\times3$ matrix with all elements 
equal to zero. 
 The full neutrino mass $\mathcal{M}$
is diagonalised by a $(3+n)\times (3+n)$ unitary matrix $U$, 
through a well known rotation between the neutrino flavour and 
mass eigenstates. We give below the relation between 
the left-handed (LH) components of the corresponding fields 
($\nu_{\ell L}$, $\nu^c_{sL}$ and $\chi_{iL}$, $N_{kL}$):
\be
\begin{pmatrix}
\nu_{\ell L} \\ \nu^c_{s L} \end{pmatrix} = U \begin{pmatrix}
\chi_{iL} \\ N_{k L} \end{pmatrix}.
\ee
%
Taking into account that the active block of $U$ is unitary 
to a very good approximation, the 
complete mixing matrix can be expanded as
\footnote{In the following we work in the basis in which the 
charged lepton mass matrix is diagonal.}
\be
U =\begin{pmatrix} 1-\theta\theta^\dagger/2 & \theta \\
 -\theta^\dagger & 1-\theta^\dagger \theta/2 \end{pmatrix} 
\begin{pmatrix} U_{\text{PMNS}} & 0\\
 0 & V \end{pmatrix} +\mathcal{O}\left( \theta^3 \right) 
 = \begin{pmatrix} U_{\text{PMNS}} & \theta V\\
 -\theta^\dagger U_{\text{PMNS}} & V \end{pmatrix} 
+\mathcal{O}\left( \theta^2 \right)
 \,,
\label{U}
\ee
%
where $\theta$ is a $3\times n$  matrix with ``small'' entries, 
which characterises the mixing between the active and 
the sterile neutrinos, $U_{\text{PMNS}}$ is the PMNS neutrino mixing matrix \cite{BPont57,MNS62} 
and $V$ is a $n\times n$ unitary matrix. 
The quantity $(\theta\,V)_{\ell k}$, $\ell=e,\mu,\tau$, $k=1,\ldots,n$,
is the coupling of the heavy 
Majorana neutrino $N_k$ to the charged lepton $\ell$ in the weak 
charged lepton current, and to the flavour neutrino $\nu_\ell$ 
in the weak neutral lepton current.

 From the diagonalization of the complete neutrino mass matrix $\mathcal{M}$, 
at leading order in $\theta$ we have \cite{Ibarra:2010xw}
\bea
  \theta^* \,M_R\, \theta^\dagger 
&\approx& 
- U_{\text{PMNS}}^*\,\hat m \,U_{\text{PMNS}}^\dagger\,,
\label{constraints1}\\  
  \theta^* \,M_R &\approx& m_D\,,
\label{constraints2}\\
 M_R &\approx& V^*\,\hat M \,V^\dagger \,,
\label{constraints3}
\eea
%
where
\begin{equation}
\label{diagmasses}
 \hat m \;\equiv\; \text{diag}(m_1,m_2,m_3)\,,\quad\quad 
\hat M \;\equiv\; \text{diag}(M_1,\ldots,M_n)\,.
 \end{equation}
%
It follows from Eqs. (\ref{constraints1}) and  (\ref{constraints3}) that
\begin{equation}
(\theta\,V)^* \,\hat M\, (\theta\,V)^\dagger 
\approx
- U_{\text{PMNS}}^*\,\hat m \,U_{\text{PMNS}}^\dagger\,.
\label{constraints4}
 \end{equation}
%
In terms of the seesaw parameters we have for 
the active-sterile neutrino mixing:
\be
\theta^* \approx m_D \,M_R^{-1}.
\label{theta}
\ee
%
Using Eqs.~(\ref{constraints1}) and (\ref{theta}), we recover the 
usual type I seesaw relation for the (tree-level) 
light neutrino mass matrix, namely
\begin{equation}
\label{mvtreetypeI}
m_{\nu}^{\rm tree}\;=\; -m_D\, M_R^{-1} \,m_D^T\; 
\equiv\; -\theta^* \,M_R\, \theta^\dagger \; 
=\; -(\theta\,V)^* \, \hat M\, (\theta\,V)^\dagger\,
=\, U_{\text{PMNS}}^*\,\hat m \,U_{\text{PMNS}}^\dagger\,.
\end{equation}
%

The effective Majorana neutrino mass, $m_{\beta\beta}$, which enters in 
the  $0\nu\beta\beta$ decay amplitude, receives, in general, two 
different contributions,
corresponding to the exchanges of the light and heavy virtual 
Majorana neutrinos: 
\begin{equation}
	m_{\beta\beta} \;=\;  m_{\beta\beta}^{\rm light}\,+\,m_{\beta\beta}^{\rm heavy} \,,\label{mefftot}
\end{equation}
%
with 
\begin{equation}
	m_{\beta\beta}^{\rm light} \;=\; 
\sum\limits_{i=1}^3\,(U_{\rm PMNS})_{ei}^2\,m_i 
\;=\; - \sum\limits_{k}\,(\theta\,V)^2_{ek}\,M_k\,,
\label{mefflight}
\end{equation}
%
where we have used Eq.~(\ref{constraints4}), which holds at tree-level in the type I seesaw models.
 A good estimate for the contribution due to the heavy Majorana 
neutrino exchange for $M_k\gg 100$ MeV is 
\cite{Blennow:2010th}
\begin{equation} 
 m_{\beta\beta}^{\rm heavy} \approx
- \sum_{k} (\theta V)_{ek}^2\, f(A)\, 
(M_{a}/M_{k})^{2}\,M_k \,,
\label{meeh1}
\end{equation}
%
where $M_{a}\approx 0.9$ GeV and  $f(A)$
depends on the decaying isotope considered. 
For, e.g.,   $^{48}$Ca, 
$^{76}$Ge, $^{82}$Se, $^{130}$Te and $^{136}$Xe, 
the function $f(A)$ takes the values $f(A)\approx$ 
0.033, 0.079, 0.073, 0.085 and 0.068, respectively.

\begin{figure}[t!]
\includegraphics[width=0.6\textwidth,angle=0]{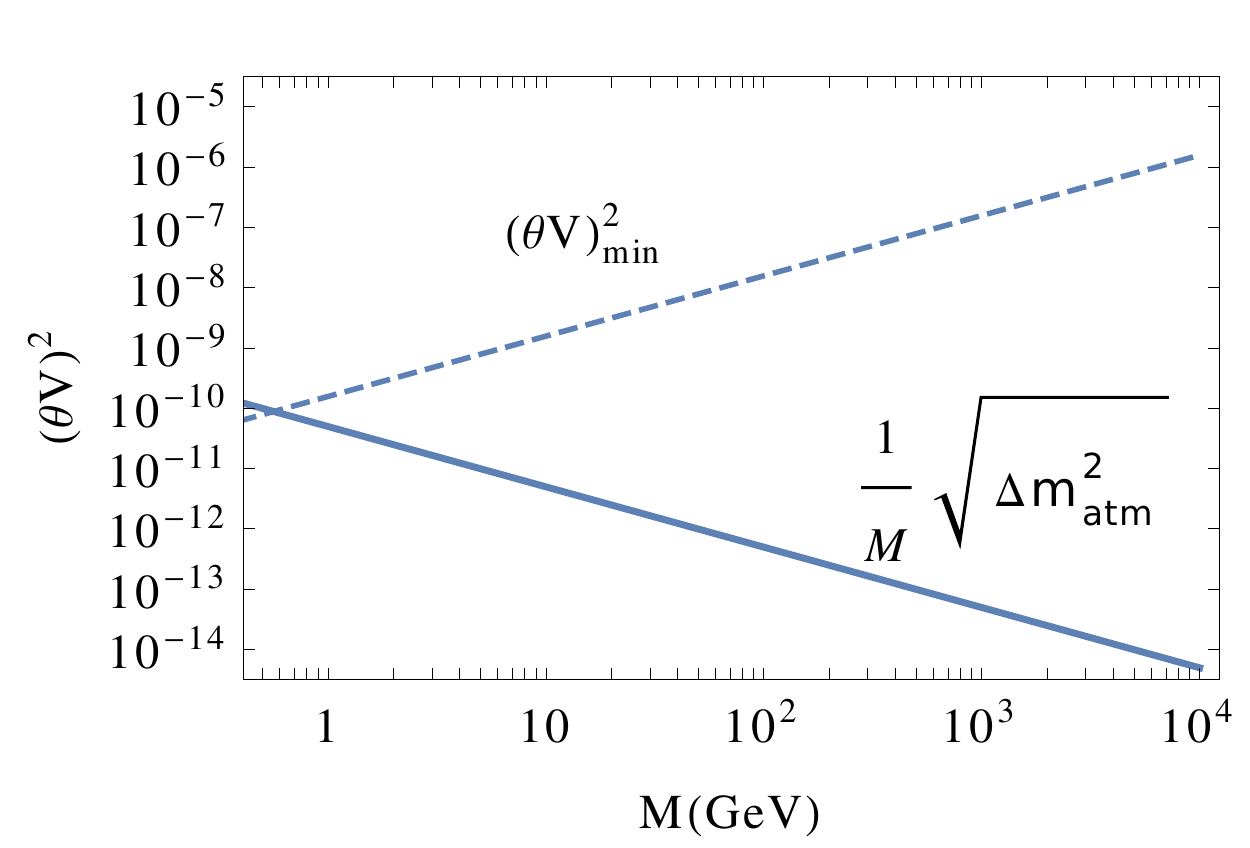} 
\caption{
\label{estimate}
{\small \textbf{Active-sterile neutrino mixing.}   The dashed line stands for 
an estimate of the minimum $(\theta V)^2$ 
required in order to have 
$|m_{\beta\beta}^{\rm heavy}| > 10^{-2}$ eV in the case of 
$0\nu\beta\beta$ decay of $^{76}$Ge. 
The solid line corresponds to the naive seesaw scaling of 
$(\theta V)^2$ (see the text for further details).}}
\end{figure}

 Using Eq. (\ref{meeh1}), 
 it is easy to estimate the minimum mixing $(\theta V)_{\rm min}$ 
required in order to have a contribution
at the aimed sensitivity of the next generation of $0\nu\beta\beta$ 
decay experiments, that is $|m_{\beta\beta}^{\rm heavy}|\gtrsim 10^{-2}$ eV. 
In Fig.~\ref{estimate} we compare this 
estimate for $(\theta V)_{\text{min}}$   for the $^{76}$Ge isotope,
$(\theta V)^2_{\text{min}} \simeq 1.6\times 10^{-10} M~\text{GeV}^{-1}$ (dashed line), 
with the naive seesaw scaling suggested by Eq.~(\ref{mvtreetypeI}),
$(\theta V)^2_{\text{naive}}=\sqrt{\Delta m^2_{\text{atm}}}/M 
\simeq 5\times 10^{-11}~\text{GeV}/M$ (solid line) 
as a function of the RH neutrino mass scale  $M$ 
(expressed in units of GeV).

From Fig.~\ref{estimate} it is clear that for RH neutrino masses 
larger than $\sim1$~GeV a considerable enhancement with respect to the naive 
seesaw scaling of $\theta V$ is required in order to have a 
sizable RH neutrino contribution. Obviously, this enhancement 
increases with the mass of the RH neutrinos. 
We notice that in the region 
$M\approx500$ MeV$\textendash 1$ GeV,  
the naively estimated mixing, $(\theta V)^2_{\text{naive}}$, 
is in the right ballpark.
Similar conclusions are valid for 
$(\theta V)^2_{\text{min}}$ and  $(\theta V)^2_{\text{naive}}$ 
in the cases of $0\nu\beta\beta$ decay of other 
isotopes ($^{48}$Ca, $^{82}$Se, $^{130}$Te, $^{136}$Xe, etc.).
%
\subsection{Casas-Ibarra Parametrization and 
Large Active-Sterile Neutrino Mixing}

In order to understand under which conditions an enhancement 
with respect to the naive scaling of the active-sterile mixing 
(or equivalently, of the charged current couplings of the 
heavy Majorana neutrinos $(\theta V)_{\ell k}$)
can be expected, we employ the Casas-Ibarra parametrization of  
$\theta V$~\cite{Casas:2001sr}. 
In this parametrization 
the light neutrino masses and the angles and phases of the PMNS 
matrix are input parameters, in such a 
way that the correct light neutrino mixing pattern is always recovered. 
The Casas-Ibarra parametrization is obtained 
rewriting Eq.~(\ref{constraints1}) as
\be
\left(\pm i \,\hat m^{-1/2} \,U_{\text{PMNS}}^\dagger \,\theta V \,\hat{M}^{1/2}\right)\,\left(\pm i \,\hat m^{-1/2}\, U_{\text{PMNS}}^\dagger \,\theta V\, \hat M^{1/2}\right)^T
\equiv R\, R^T=1\,,
\ee
where $R$ is a general $3 \times n$ complex  
matrix which parametrizes the new physics
degrees of freedom associated to the sterile neutrino sector. 
Using this parametrization, $\theta V$ can be written as
\be
\theta V  = 
\mp\, i\,  U_{\text{PMNS}}\, \hat m^{1/2}\, R\, \hat M^{-1/2}\,.
\label{thetaR}
\ee
%
The matrix  $V$ can be set to the unit matrix if one works in the basis 
in which the Majorana sub-matrix $M_R$ is diagonal.~\footnote{ 
An extension of this parametrization to all orders in the seesaw 
expansion can be found in~\cite{Donini:2012tt,Blennow:2011vn}.}

Naively, from Eq.~(\ref{mvtreetypeI}) 
one may conclude that 
$\theta V\approx\mathcal{O}\left(\sqrt{\frac{\hat m}{\hat M}}\right)$, 
i.e., that the mixing (or coupling) $\theta V$ is expected to be suppressed by
the heavy neutrino mass scale. However, having a larger mixing 
is perfectly possible
due to an enhancement factor contained  in the matrix $R$ 
\cite{Ibarra:2010xw,Ibarra:2011xn}. Obviously, such enhancement can only 
be in agreement with the light 
neutrino spectrum if there is a non-trivial suppression/cancellation 
in the l.h.s.~of Eq.~(\ref{constraints4}). This 
extra suppression is related to particular textures of 
the neutrino mass matrix, which can be motivated,
for instance, introducing an extra $U(1)$ global symmetry 
in the Lagrangian, as it is the case in the 
so called ``inverse'' and ``direct'' seesaw models
~\cite{Mohapatra:1986bd,Branco:1988ex}.
In these models the indicated
global symmetry can be identified with that 
corresponding to the conservation of 
 a non-standard lepton charge (see further). 

In the following we will focus on the minimal seesaw scenario 
with $n=2$ RH sterile neutrinos~\footnote{In the present article we will use the term 
``heavy Majorana neutrinos'' for Majorana neutrinos having masses 
exceeding approximately 100 MeV.}~(see, e.g., \cite{3X2Models}) giving rise 
to two heavy Majorana mass-eigenstate neutrinos,
which predicts one massless  and 
two massive light active neutrinos. For the light neutrino mass spectrum with 
normal hierarchy (NH) and inverted hierarchy (IH) we have 
\begin{eqnarray}
\label{measuredmassesNH}
&& m_1=0 \,, \quad \quad  m_2= \sqrt{\Delta m^2_{21}} \,, 
\quad \quad m_3= \sqrt{\Delta m^2_{31}} \, ,
\quad\quad\mbox{(NH)}\\
&& m_1= \sqrt{ |\Delta m^2_{32}| - \Delta m^2_{21} } \,, \quad\quad m_2= \sqrt{|\Delta m^2_{32}| } \,, 
\quad\quad   m_3= 0\quad \quad\mbox{(IH)} \,.
\label{measuredmassesIH}
\end{eqnarray}
%
The current best fit values obtained from the global fit analysis in 
\cite{nufit} are 
\begin{eqnarray}
\label{dmbf}
&\Delta m^2_{21} = 7.50 \times 10^{-5} \; \mathrm{eV}^2 \; ,&\\ 
\nonumber
&\Delta m^2_{31} = 2.457 \times 10^{-3} \;\; \mathrm{eV}^2  \;\; \mbox{(NH)} \;\;\; 
\mbox{and} \;\;\;  
\Delta m^2_{32} =-2.449 \times 10^{-3} \;\mathrm{eV}^2 \;\; \mbox{(IH)} \; .
&
\end{eqnarray}
%
In this minimal seesaw scenario, the two (tree-level) 
non-zero light neutrino masses $m_2^{\rm tree}$ and $m_3^{\rm tree}$ 
($m_1^{\rm tree}$) in the case of NH (IH) neutrino mass spectrum 
satisfy the relation:
 \begin{equation}
\label{treeproduct}
 m_{2}^{\rm tree}\,m_{3(1)}^{\rm tree}\;
\equiv\; - \det[M_{R}^{-1}]\det[m_{D}^{T}m_{D}]\,,~~~{\rm NH~(IH)}\,,
 \end{equation}
%
 which is basis independent.

In the considered case the $R$-matrix, which enters into Eq. (\ref{thetaR}), 
can be parametrized~as~\cite{Ibarra:2011xn}
\bea
R &= &\begin{pmatrix} 0 & 0 \\
\cos\left(\theta_{45}+i\gamma\right) & -\sin\left(\theta_{45}+i\gamma\right) \\
 \sin\left(\theta_{45}+i\gamma\right) & \cos\left(\theta_{45}+i\gamma\right) \end{pmatrix}\,,\quad\quad\text{for \:\:NH}\,,
\label{RNO}\\
R &= &\begin{pmatrix} 
\cos\left(\theta_{45}+i\gamma\right) & -\sin\left(\theta_{45}+i\gamma\right) \\
 \sin\left(\theta_{45}+i\gamma\right) & \cos\left(\theta_{45}+i\gamma\right) \\
 0 & 0  \end{pmatrix}\,,\quad\quad\text{for \:\:IH}\,,
\label{RIO}\
\eea
%
where $\theta_{45}$ and $\gamma$ are real 
parameters.
If $R$ were real, i.e., $\gamma=0$, there is no way to obtain 
any enhancement of the couplings/mixings $\theta V$ of interest 
since $R$ would essentially be a 
real orthogonal matrix. However, 
for $\gamma\neq0$ and $e^{\pm\gamma}\gg 1$ 
an enhancement of $\theta V$  is possible:
\bea
|\cos\left(\theta_{45}+i\gamma\right)|^2 &=& \cos^2\theta_{45}+\sinh^2\gamma\gg 1\Leftrightarrow e^{\pm\gamma}\gg1\,,
\nonumber\\
|\sin\left(\theta_{45}+i\gamma\right)|^2 &=& \sin^2\theta_{45}+\sinh^2\gamma\gg 1\Leftrightarrow e^{\pm\gamma}\gg1\,.
\eea
%
In fact, for $e^{\pm\gamma}\gg 1$ the expression of $R$ 
in the NH case 
reduces to
\be
R \approx e^{-i\,\theta_{45}}\,\frac{e^{\pm\gamma}}{2}\,
\begin{pmatrix} 0&0 \\ 1 & \pm i \\
 \mp i & 1 \end{pmatrix}\,,~~~{\rm NH}\,.
\label{RissNH}
\ee
%
Similarly, one can derive from (\ref{RIO}) the same limit of $R$ for 
the IH neutrino mass spectrum:
\be
R \approx e^{-i\,\theta_{45}}\,\frac{e^{\pm\gamma}}{2}\,
\begin{pmatrix} 1 & \pm i \\
 \mp i & 1 
\\  0 & 0 
\end{pmatrix}\,,~~~{\rm IH}\,.
\label{RissIH}
\ee
%
Notice that the Casas-Ibarra parameter $\gamma$  in (\ref{RissNH}) and (\ref{RissIH}) can be related to the maximum eigenvalue $y$ \cite{Ibarra:2011xn} of the Dirac mass matrix $m_D$ in Eq.~(\ref{Mnu}), that is 
\begin{eqnarray}
	y^2\,v^2 & = & 2\, \text{max}\left\{ \text{eig}\left( m_D\, m_D^\dagger \right) \right\}\;=\;\frac12 e^{\pm \gamma}\,M_1\left(m_2+m_3 \right)\left(2+z\right)\,,\quad \quad \text{NH}\,,\\
	y^2\,v^2 & = & 2\, \text{max}\left\{ \text{eig}\left( m_D\, m_D^\dagger \right) \right\}\;=\;\frac12 e^{\pm \gamma}\,M_1\left(m_1+m_2 \right)\left(2+z\right)\,,\quad \quad \text{IH}\,,
\end{eqnarray}
where $z$ denotes the relative mass splitting of the two heavy Majorana neutrino masses, $z=(M_2-M_1)/M_1$, and $v=246$ GeV is the EW symmetry breaking scale. 

Introducing the expression (\ref{RissNH}) (or  (\ref{RissIH}))
in Eq.~(\ref{thetaR}) 
one obtains \cite{del Aguila:2006dx,Gavela:2009cd, Ibarra:2010xw,Ibarra:2011xn} 
\be
\frac{\left(\theta V\right)_{\ell 1}}{\left(\theta V\right)_{\ell 2}} \approx\pm i\,\sqrt{\frac{M_2}{M_1}}.
\label{condition}
\ee
%
Then, in terms of $y$ the active-sterile neutrino mixing in Eq.~(\ref{thetaR})
takes the form \cite{Ibarra:2011xn}
\begin{eqnarray}
\label{mixing-vs-y}
\left|\left(\theta V\right)_{\ell 1} \right|^{2}&=&
\frac{1}{2\,(2+z)}\frac{y^{2} v^{2}}{M_{1}^{2}}\frac{m_{3}}{m_{2}+m_{3}}
	\left|U_{\ell 3}+i\sqrt{m_{2}/m_{3}}\,U_{\ell 2} \right|^{2}\,,
\quad{\rm NH}\,,\\
\left|\left(\theta V\right)_{\ell 1} \right|^{2}&=&
\frac{1}{2\,(2+z)}\frac{y^{2} v^{2}}{M_{1}^{2}}\frac{m_{2}}{m_{1}+m_{2}}
	\left|U_{\ell 2}+i\sqrt{m_{1}/m_{2}}\,U_{\ell 1} \right|^{2}\,,
\quad {\rm IH}\,.
\label{mixing-vs-yIH}
\end{eqnarray}
All in all, the previous relations imply that in the basis 
in which the RH neutrino Majorana mass term is diagonal,
the neutrino Yukawa couplings, or equivalently 
$(m_D)_{\ell 1}$ and $(m_D)_{\ell 2}$,
should satisfy the following relation:
\be
\frac{\left(m_D\right)_{\ell 1}}{\left(m_D\right)_{\ell 2}}\approx 
\pm i\,\sqrt{\frac{M_1}{M_2}}
\label{conditionMD}
\ee
%
Any texture of the neutrino mass matrix which satisfies 
this condition gives rise to relatively large couplings $\theta V$ 
with the right suppression/cancellation in the light (flavour) neutrino
mass matrix, which allows to recover the correct light neutrino mass 
spectrum at tree-level. 
The relatively large  $\theta V$ 
thus generated can 
saturate the present bounds even 
in the case in which the heavy Majorana neutrino 
spectrum is hierarchical. 

Using Eqs.~(\ref{meeh1}) and (\ref{condition}), 
one can easily estimate the contribution 
to the $0\nu\beta\beta$ decay effective Majorana mass due to the 
exchange of the heavy Majorana neutrinos in the large coupling/mixing 
case of interest \cite{Ibarra:2010xw}:
%
\begin{equation} 
 m_{\beta\beta}^{\rm heavy} \approx
- \, (\theta V)_{e1}^2\, f(A)\,\frac{M^2_a}{M_1}\,
\left\{1-\left(\frac{M_1}{M_1+\Delta M}\right)^2\right\}\,,
\label{meeh2}
\end{equation}
%
with~\footnote{
 Note that $(\theta V)_{e1}^2$ depends, in particular, 
on the phase $\theta_{45}$. This implies that 
$m_{\beta\beta}$ will also depend on $\theta_{45}$~\cite{Ibarra:2011xn}.}~$\Delta M =M_2-M_1$. 
Clearly, if $\Delta M \ll M_1$ the contribution will be proportional 
to $\Delta M$, while in the limit $\Delta M \gg M_1$ the dependence 
on $\Delta M$ is subleading since the lightest 
RH neutrino dominates the contribution.

The interplay between the light and heavy Majorana neutrino exchange 
contributions in the effective Majorana mass,
$m_{\beta\beta} =  m_{\beta\beta}^{\rm light} + m_{\beta\beta}^{\rm heavy}$, 
in the scheme under discussion in which  
Eq. (\ref{condition}) holds and  $m_{\beta\beta}^{\rm heavy}$ is 
given by Eq. (\ref{meeh2}),  was investigated in detail 
in \cite{Ibarra:2011xn} in the case when the two heavy Majorana 
neutrinos form a pseudo-Dirac pair,
$0 < \Delta M =M_2-M_1 \ll M_1,M_2$, and have masses in the interval 
$\sim (50 - 1000)$ GeV. It was found that there exists a 
relatively large region of the allowed parameter space of the scheme 
in which the heavy Majorana neutrino contribution 
can change drastically the predictions based on the light Majorana 
neutrino exchange contribution. More specifically, it was found 
 that  \cite{Ibarra:2011xn}:
i) $|m_{\beta\beta}|$ in the case of NH spectrum 
can have values in the interval 
$0.01~{\rm eV}\lesssim |m_{\beta\beta}| \lesssim 0.1$ eV, i.e., 
in the range of sensitivity of the current 
GERDA \cite{GERDA}, EXO \cite{EXO200}, Kamland-Zen \cite{KamlandZen} and CUORE \cite{CUORE0} experiments 
and of a few other experiments under preparation 
(Majorana  \cite{Abgrall:2013rze}, SNO+ \cite{Hartnell:2012qd}, AMORE \cite{Bhang:2012gn}, etc.). We recall that in the case of 
$0\nu\beta\beta$ decay generated only by 
light Majorana neutrino exchange we have 
(see, e.g., \cite{bb0nuNH2008,PDG2014}) 
$|m_{\beta\beta}| =  |m_{\beta\beta}^{\rm light}| \lesssim 0.005$ eV;\\ 
ii) $|m_{\beta\beta}|$ in the case of IH spectrum  
can be strongly suppressed due to partial, or even total, 
cancellation between $m_{\beta\beta}^{\rm light}$ and 
$m_{\beta\beta}^{\rm heavy}$ in $m_{\beta\beta}$ (see also \cite{Pascoli:2013fiz}). 
Since the magnitude of $m_{\beta\beta}^{\rm heavy}$, as it follows from 
Eq.~(\ref{meeh2}), depends on the atomic number $A$ of the 
decaying nucleus \cite{HPR83}, 
the cancellation between $m_{\beta\beta}^{\rm light}$ and 
$m_{\beta\beta}^{\rm heavy}$ in $m_{\beta\beta}$ can take place 
for a given nucleus (say, e.g., for $^{48}$Ca)
but will not hold for other 
nuclei ($^{76}$Ge, $^{82}$Se, $^{130}$Te, $^{136}$Xe, etc.).
If the $0\nu\beta\beta$ decay is due only to the  
light Majorana neutrino exchange we have 
in the case of IH spectrum, as is well known \cite{bb0nuNHIH} 
(see also, e.g., \cite{PDG2014}), 
$ 0.013~{\rm eV} \lesssim |m_{\beta\beta}| =  
|m_{\beta\beta}^{\rm light}| \lesssim 0.050$ eV.

On the other hand, in \cite{Ibarra:2011xn} the role of the one-loop corrections was not studied. In 
\cite{LopezPavon:2012zg} it was shown that the one-loop corrections to the light neutrino 
masses generated in the scheme under discussion turn out to be very relevant. Essentially,
a sizable heavy contribution to the $0\nu\beta\beta$ decay for heavy masses in the 
range $\sim (50 - 1000)$ GeV generates at the same time a very large one-loop correction
to the light neutrino masses. In this work we analyse in detail the role of the
one-loop effects showing that similar conclusions to the ones drawn in \cite{Ibarra:2011xn}
will be obtained. However, we will also show that the price one has to pay in order to
have a significant impact of the heavy neutrinos in the $0\nu\beta\beta$ decay is the 
requirement of a highly fine-tuned cancellation between the tree-level and one-loop  contributions
to the light neutrino masses.

\subsection{Comparison with Extended and 
Inverse Seesaw Scenarios}

As an application of the previous results, 
we consider the effect of heavy RH neutrinos on the 
$0\nu\beta\beta$ decay amplitude in the case of two 
different realisations of the type I seesaw scenario,
which predict  a large active-sterile neutrino mixing $\theta V$, that is the 
well known extended seesaw (ESS)~\cite{Kang:2006sn}  and 
inverse/direct seesaw (ISS)~\cite{Mohapatra:1986bd, Branco:1988ex} models. 
In particular, we will clarify how the large mixing realisations 
described in the previous section in terms of the Casas-Ibarra 
parametrization match with the ISS and ESS scenarios.
 
 In order to understand the predictions in these classes of models 
it is useful to  adopt the following parametrization of the generic 
mass terms in the seesaw Lagrangian (\ref{typeI}), namely
\begin{eqnarray}
 \mathcal{M}\;\equiv\; \left( \begin{array}{cc} \mathbf{O} &  m_D \\ m_D^T  
& M_R 
 \end{array} \right)\; =\; \left( \begin{array}{ccc} \mathbf{O} &  \mathbf{Y}_1 \,v/\sqrt{2}& \epsilon \,\mathbf{Y}_2 \,v/\sqrt{2} \\  \mathbf{Y}_1^T v/\sqrt{2}  
&  \mu' & \Lambda \\ \epsilon \,\mathbf{Y}_2^T \,v/\sqrt{2}
 & \Lambda & \mu  
 \end{array} \right)\,, 
 \label{Mnu2}
\end{eqnarray}
%
where
$\mathbf{Y}_i\equiv(y_{i e},y_{i \mu}, y_{i \tau})^T$, for  $i=1,2$. This parametrization is completely general and, in principle, 
$\epsilon$, $\mu$, $\mu'$ and  $\Lambda$ 
can take any value.~\footnote{In the following we will assume for 
simplicity that all the parameters introduced in Eq.~(\ref{Mnu2}) are real.}~However, 
$\epsilon$, $\mu$ and $\mu'$ can be interpreted as lepton 
number violating couplings and, therefore, in principle they take arbitrarily 
small values, 
because in this case there is an approximate global 
symmetry of the seesaw Lagrangian corresponding to the 
conservation of the lepton charge $L' = L_e + L_{\mu} + L_{\tau} + L_1 - L_2$, 
where $L_1$ and $L_2$ are the charges carried by the 
RH neutrino fields $\nu_{1R}$ and $\nu_{2R}$, respectively.
In the limit of  $\epsilon = \mu = \mu' = 0$, the conservation of $L'$ 
is exact. In this case the neutrino sector consists of three massless neutrinos
and one massive Dirac fermion, which can be inferred, in particular,
directly from the expression of the charge $L'$ in terms of the charges 
$L_{\ell}$ and $L_{1,2}$ \cite{Leung:1983ti,Bilenky:1987ty}. The exact conservation of $L'$ 
corresponds to the case in which condition (\ref{condition}) is exactly fulfilled and 
the RH neutrino splitting satisfies: $\Delta M =M_2-M_1 \rightarrow 0$.

 In terms of the new parameters, the exact (tree-level) expression 
of the light neutrino mass matrix given in (\ref{mvtreetypeI}) 
is proportional to $\mu$ and $\epsilon$, that is
\be
\label{treemass}
m_{\nu}^{\rm tree}\;=\;\frac{v^2}{2\,(\Lambda^2-\mu'\mu)}
\left(\mu \,\mathbf{Y}_1\,\mathbf{Y}_{1}^T\,+\, \epsilon^2\,\mu' \,\mathbf{Y}_2\, \mathbf{Y}_{2}^T\,-\,\Lambda\,\epsilon\, (\mathbf{Y}_2\, \mathbf{Y}_{1}^T+\mathbf{Y}_1\, \mathbf{Y}_{2}^T)\right)\,,
\ee
%
and thus if $\mu=\epsilon=0$ there is a complete 
cancellation at tree-level for the light neutrino masses. 
As we will see in the next section, if $\mu'$ is different from zero, 
at least one neutrino mass can be generated at one-loop, even for 
$\mu=\epsilon=0$~\cite{LopezPavon:2012zg}. Furthermore, from the 
diagonalization of (\ref{treemass}),  we obtain for the product of the smallest ($m_l^\text{tree}$) and 
the largest ($m_h^{\rm tree}$) light neutrino masses:
\begin{eqnarray}
	&& \left| m_{l}^{\rm tree}\,m_{h}^{\rm tree} \right| \;=\;\left| \det \left[M_R^{-1}\right]\,\det\left[m_D^T\,m_D\right]\right|\;=\nonumber\\\\
	&& \frac{v^{4}\,\epsilon^{2}\left| y_{2 e}^{2}\,(y_{1 \mu}^{2}+y_{1 \tau}^{2})+
	 y_{1 e}^{2}(y_{2 \mu}^{2}+y_{2 \tau}^{2}) -2\, y_{1 e} y_{2 e} (y_{1 \mu} y_{2\mu}+y_{1 \tau}y_{2 \tau})+(y_{2 \mu}\, y_{1 \tau}-y_{1 \mu}y_{2\tau})^{2} \right| }{4|\Lambda^{2}-\mu\,\mu^{\prime}|}\,.
\nonumber
\end{eqnarray}
%
 From this relation it follows that in order to have 
two massive active neutrinos at tree-level, i.e., 
$m_{l,h}^{\rm tree}\neq 0$, $i)$ an explicit breaking of the 
lepton charge conservation via the neutrino Yukawa 
couplings is necessary, that is the parameter $\epsilon$ must always be 
 different from zero; 
$ii)$ the vectors of neutrino Yukawa couplings $\mathbf{Y}_{1}$ and $\mathbf{Y}_{2}$
cannot be proportional. 

Accordingly,  the two seesaw limits of Eq.~(\ref{treemass}) which give rise to
 large active-sterile neutrino mixing $\theta V$ and generate 
sufficiently small active  
neutrino masses are:
\begin{itemize}
 \item i)  $\mu'\gg \Lambda,\,y_{1 \alpha}\,v \gg \mu,\,\epsilon\, y_{2 \alpha}\, v$ (\textbf{ESS limit}). 
This limit matches the so-called extended seesaw~\cite{Kang:2006sn} 
models and corresponds to a hierarchical spectrum for the heavy neutrinos:
\begin{eqnarray}
M_1 &\approx& (\Lambda^2/\mu'-\mu)\,,\quad\;\;
(\theta V)_{\ell 1}\;\approx\; i\frac{v}{\sqrt{2}\,M_1}
\left [ y_{1\ell}\,\frac{\Lambda}{\mu' - \mu} 
- \epsilon \, y_{2\ell} \left (1 - \frac{\Lambda^2}{2(\mu' - \mu)^2}\right)
\right]\,,\nonumber\\&&
\label{ESS}\\
M_2 &\approx&\mu'\,+\,\Lambda^2/\mu'\,,\quad\;\; 
(\theta V)_{\ell 2}\;\approx\; 
\frac{v}{\sqrt{2}\, M_2}
\left [ y_{1\ell}\, \left( 1 - \frac{\Lambda^2}{2(\mu' - \mu)^2}\right ) 
+\, \epsilon \, y_{2\ell} \, \frac{\Lambda}{\mu' - \mu}\,
\right ]\,,\nonumber\\ &&
\label{ESSb}
\end{eqnarray}
where we also show the corresponding mixing with the 
active neutrinos. Then, the approximate tree-level contribution 
to the $0\nu\beta\beta$ decay effective Majorana 
mass due to the exchange of the light and the heavy neutrinos is 
\bea
\label{ESSbb0nul}
m^{\text{light}}_{\beta\beta} &\approx& \frac{v^2}{2\left(\Lambda^2/\mu'-\mu\right)}\left(\frac{\mu}{\mu'}\,y_{1e}^2\, -\,2\,\epsilon\frac{\Lambda}{\mu'}\,y_{1e}\,y_{2e}
\right)\,,\\
m^{\text{heavy}}_{\beta\beta} &\approx& 
f(A)\,\frac{v^2\, M^2_a}{2\left(\Lambda^2/\mu'-\mu\right)^3}
\left(\frac{\Lambda^2}{\mu'^2}\,y_{1e}^2\, 
 -\,2\,\epsilon\,\frac{\Lambda}{\mu'}\,y_{1e}\,y_{2e}\right)\,,
\label{ESSbb0nuh}
\eea
%
respectively. 
The dominant term in $m^{\text{heavy}}_{\beta\beta}$
is due to the exchange of the lighter of the two heavy Majorana 
neutrinos $N_1$, the exchange of $N_2$ giving a subleading 
(and negligible in the leading approximation we employed)
correction.
Notice that, 
if $\Lambda^2/\mu'\gg\mu$, $m^{\text{light}}_{\beta\beta}$ 
becomes independent of $\mu'$ while 
$m^{\text{heavy}}_{\beta\beta}$ is proportional to $\mu'$:
\bea
\label{mbbESS1}
m^{\text{light}}_{\beta\beta} &\approx& \frac{v^2}{2\,\Lambda^2}\left(\mu \,y_{1e}^2 \,-\,2\,\epsilon\,\Lambda\, y_{1e}\,y_{2e}\right)\,,
\\
m^{\text{heavy}}_{\beta\beta} &\approx& 
f(A)\,\frac{\mu' \, v^2\, M^2_a}{2\,\Lambda^4}
\left(\,y_{1e}^2 \,
-\,2\,\epsilon\,\frac{\mu'}{\Lambda}\,y_{1e}\,y_{2e}\right)\,.
\label{mbbESS2}
\eea

\item ii) $\Lambda\gg y_{1\alpha}\,v \gg \mu',\mu,\,\epsilon\, y_{2\alpha}\, v$ 
(\textbf{ISS limit}). This limit corresponds to a minimal realisation with
only two RH neutrinos of the so-called inverse or direct 
seesaw models~\cite{Gavela:2009cd}. 
In this case the heavy neutrino spectrum is quasi-degenerate, 
forming a quasi-Dirac pair \cite{Wolfenstein:1981kw, Petcov:1982ya}
\begin{eqnarray}
M_1 &\approx&\Lambda -\frac{\mu+\mu'}{2}\,,\quad\;\; 
(\theta V)_{\ell 1}\,\approx\;
i\frac{v}{2M_1}\,
\left[y_{1\ell}\left(1 + \frac{\mu - \mu'}{4\Lambda}\right) 
-\,\epsilon \,y_{2 \ell}\left(1 - \frac{\mu - \mu'}{4\Lambda}\right)  
\right]\,,\nonumber\\ &&
\label{ISS}\\
M_2 &\approx& \Lambda\,+\,\frac{\mu\,+\,\mu'}{2}\,,\quad 
(\theta V)_{\ell 2} \;\approx\; 
\frac{v}{2M_2}\,\left[y_{1 \ell}\left(1 - \frac{\mu - \mu'}{4\Lambda}\right) 
+\,\epsilon \,y_{2 \ell}\left(1 + \frac{\mu - \mu'}{4\Lambda}\right)  
\right]\,,\nonumber\\ &&
\label{ISSb}
\end{eqnarray}
In this limit the light and heavy contributions to 
the $0\nu\beta\beta$ decay rate are given by:
\bea
 \label{mbbISSl}
m^{\text{light}}_{\beta\beta} &\approx& 
\frac{v^2}{2\,\Lambda^2}\left(\mu\, y_{1e}^2 
- 2\,\epsilon\,\Lambda \,y_{1e}\,y_{2e}\right)\,,\\
m^{\text{heavy}}_{\beta\beta} &\approx& 
f(A)\,\frac{v^2\, M^2_a}{2\,\Lambda^4}
\left((2\,\mu\,+\,\mu')\, y_{1e}^2 \,
-\,2\,\epsilon\,\Lambda\, y_{1e}\,y_{2e}\right)\,.
\label{mbbISS}
\eea
Both of them are proportional to the 
small lepton number violating parameters, as it should be. 
Notice that the expression
of $m^{\text{light}}_{\beta\beta}$ above is exactly the same 
as the one given in Eq.~(\ref{mbbESS1}). 
\end{itemize}

On one hand, it follows from 
Eqs.~(\ref{mbbESS1}), (\ref{mbbESS2}), (\ref{mbbISSl}) and 
(\ref{mbbISS}) that a relatively large contribution to the 
$0\nu\beta\beta$ decay rate due to the heavy Majorana neutrino exchange 
might be possible at tree-level without affecting the smallness of the
light neutrino masses since in the limits considered here
$m^{\text{heavy}}_{\beta\beta}\propto \mu'$,
while $m^{\text{light}}_{\beta\beta}$ is independent of $\mu'$.
On the other hand, Eqs.~(\ref{ESS}-\ref{ESSb}) and (\ref{ISS}-\ref{ISSb}) 
confirm that the condition to obtain relatively large mixings, 
Eq.~(\ref{condition}), is fulfilled at leading order, 
that is in the Casas-Ibarra 
parametrization the $R$-matrix corresponding to these two cases 
is similar to the textures reported in Eqs. (\ref{RissNH}) and 
(\ref{RissIH}).

  Finally, we  note  that in the case of the ISS model, the 
smallness of the light neutrino masses comes
from the existence of an approximate 
symmetry corresponding to the conservation of the 
lepton charge $L'$.
In contrast, in the ESS limit, the conservation of 
the lepton charge $L'$ 
is strongly violated through the $\mu'$ coupling. 
 This means that, in principle,
the one-loop corrections to the neutrino masses 
can be expected to be more important 
in the ESS limit than in the ISS one since 
in the ESS case there is no symmetry protecting the light neutrino masses
from getting relatively large corrections~\cite{LopezPavon:2012zg}.

%
\section{One-loop Corrections to the Neutrino 
Mass Matrix}\label{sec3}
%
%
We turn now to the computation of the one-loop corrections to 
the light neutrino mass matrix and the effective Majorana neutrino 
mass associated to $0\nu\beta\beta$ decay amplitude.

At one-loop the neutrino self-energy $\Sigma(p)$ provides the 
dominant finite correction to $m_\nu$~\cite{Pilaftsis:1991ug,Grimus:2002nk,AristizabalSierra:2011mn,LopezPavon:2012zg, Dev:2012sg}, 
which depends on the square of the neutrino Yukawa couplings,
as in the tree-level contribution (\ref{mvtreetypeI}), and is 
further suppressed by the one-loop factor $1/(16\,\pi^{2})$. 
In a generic basis, with the 
Dirac and Majorana mass terms defined in Lagrangian (\ref{typeI}), we obtain:
\be
 \mathcal{M}\; = \; \begin{pmatrix} m_{\nu}^{1-\text{loop}} &  m_D \\
 m_D^T & M_R  \end{pmatrix} = U^* \,\text{diag}\left(m_i, M_k\right)U^\dagger\,,
\label{Mnuloop}
\ee
%
where the new Majorana mass term generated at one-loop is in this case
\begin{equation}
 	m_{\nu}^{\rm 1-loop}\;=\; 
	\frac{1}{(4 \,\pi\, v)^2}\,m_{D}\,\left(M_{R}^{-1}\,F(M_{R}M_{R}^{\dagger})+F(M_{R}^{\dagger}M_{R})\,M_{R}^{-1} \right) \,m_{D}^{ T}\,.
\label{mv1loopB}
 \end{equation}
%
The loop function $F(x)$ is defined as
 \begin{equation}
 F(x)\equiv \frac{x}{2}\,\left(\,3\log(x/M_{Z}^{2})\,(x/M_{Z}^{2}-1)^{-1}+\log(x/M_{H}^{2})\,(x/M_{H}^{2}-1)^{-1}\,\right)\,,
\label{1loopfunc}
 \end{equation}
%
$M_H$ and $M_Z$ denoting the Higgs and the $Z$ boson mass, respectively. Hence,
the  overall light neutrino mass matrix, $m_\nu$, is given by 
the sum of the tree-level (\ref{mvtreetypeI}) and 
one-loop (\ref{mv1loopB}) terms, which 
in the basis of charged lepton mass eigenstates satisfies the relation
\begin{equation}
m_{\nu}=m_{\nu}^{\rm tree}+m_{\nu}^{\rm 1-loop}\;=\; U_{\rm PMNS}^*\,\text{diag}(m_1,m_2,m_3)\,U_{\rm PMNS}^\dagger\,.
\label{mvtot}
\end{equation}
%
The finite radiative correction given in (\ref{mv1loopB}) is in general
subdominant in the case of RH neutrinos with a high mass scale $M\gg v$, 
but it may be sizable and comparable to the tree-level term 
in seesaw scenarios where
the lepton number violating scale is taken below the TeV range. 
It is therefore interesting to analyse in greater detail the dependence 
of the light neutrino masses on the additional 
finite one-loop contribution, Eq.~(\ref{mv1loopB}).  

 In the basis in which the RH neutrino mass is diagonal, 
 the one-loop correction of interest has the following form:
\begin{eqnarray}
(m^{\rm 1-loop}_{\nu})_{\ell\ell'} 
&=&\, \frac{1 }{(4\, \pi\, v)^2}\,
(\theta V)^*_{\ell k}\,  M^3_k\,
	\left(\frac{3 \log( M_k^{2}/M_{Z}^{2})}{M_k^{2}/M_{Z}^{2}-1}
+\frac{\log(M_k^{2}/M_{H}^{2})}{M_k^{2}/M_{H}^{2}-1} \right)\,
(\theta V)^{\dagger}_{k\ell'}\,,
\label{mv1loopcorr0}
\end{eqnarray}
%
where we have used Eqs.~(\ref{constraints2}) and (\ref{constraints3}). 
The contribution of the one-loop correction 
under discussion to the effective Majorana neutrino mass 
$m^{\text{light}}_{\beta\beta}$, generated by the light Majorana 
neutrino exchange, as can be shown, is given by 
\begin{equation}\label{mbb1loopA}
	m^{\rm 1-loop}_{\beta\beta} \;=\; (m^{\rm 1-loop}_{\nu})^*_{ee}\,.
\end{equation}

%
\subsection{The Scheme with Two RH Neutrinos}
%
%

In the phenomenologically interesting scheme with two RH neutrinos, 
for each  non-zero eigenvalue $m_k$ of Eq.~(\ref{mvtot}), 
we have the exact relation
\begin{eqnarray}
	0 &=& \det\left[\, m_k\,\mathbf{1_{3\times 3}}\,+\,m_{D}\,M_{R}^{-1}\,\left(\,\mathbf{1_{2\times 2}}-\mathcal{H}(M_{R})\,\right)\,m_{D}^{T} \,\right]\nonumber\\
	&=& m_k\, \text{\rm det}\left[\, m_k\, \mathbf{1_{2\times 2}} \,+\, M_{R}^{-1}\left(\, \mathbf{1_{2\times 2}}- \mathcal{H}(M_{R})  \,\right)\,m_{D}^{T}m_{D}\,\right]\,, 
\label{id1}
\end{eqnarray} 
%
where the second equality follows form the Sylvester's determinant theorem 
and we have introduced the function~\footnote{The definition given in Eq.~(\ref{Hfun}) is by construction basis 
independent.}
\begin{equation}
	\mathcal{H}(M_{R})\;\equiv\;\frac{1}{(4\,\pi\,v)^2}\, \left(F(M_{R}M_{R}^{\dagger})+M_{R}\,F(M_{R}^{\dagger}M_{R})\,M_{R}^{-1}\right)\,.
\label{Hfun}
\end{equation} 
%
Using (\ref{id1}) and  (\ref{treeproduct}), we get the identity
\begin{eqnarray}
	\det\left[ \, \mathbf{1_{2\times 2}}- \mathcal{H}(M_{R})  \,\right]\, \left| m_{l}^{\rm tree}\,m_{h}^{\rm tree} \right| \;=\;m_l\,m_h\,, 
\label{id2}
\end{eqnarray} 
%
where $m_l$  ($m_h$) is the smaller (larger) non-zero active neutrino
mass, whose experimental value in the cases of NH and IH neutrino mass spectrum 
is given in Eqs.~(\ref{measuredmassesNH}) and (\ref{measuredmassesIH}), 
respectively.~\footnote{In the convention we are using 
$m_{l}^{\rm tree} m_{h}^{\rm tree} =m_{2}^{\rm tree} m_{3}^{\rm tree}$ 
( $m_{l}^{\rm tree} m_{h}^{\rm tree} =m_{1}^{\rm tree} m_{2}^{\rm tree}$)
and $m_l m_h = m_2 m_{3}$ ($m_l m_h = m_1 m_{2}$) in the NH (IH) case.}~Therefore, 
the determinant on the left hand side of 
Eq.~(\ref{id2}) provides a measurement of the deviation of 
the tree-level mass eigenvalues from the 
observed neutrino masses. Notice that, this is a positive 
quantity smaller than one
in the scenarios considered here. 
As a consequence of Eq.~(\ref{id2}), one has that in 
the case $m_l^{\rm tree}=0$, i.e. if two of the active neutrinos are 
massless at tree-level, it is not possible to generate at one-loop 
level two non-zero light (active) neutrino masses in the spectrum. 
In other words, in such a scenario both the solar and atmospheric 
neutrino oscillation mass differences cannot be radiatively generated.

As it is not difficult to show, 
in the minimal scenario with only two 
heavy Majorana neutrinos, in which  
condition (\ref{condition}) is exactly fulfilled, 
the  one-loop contribution to the  light neutrino mass 
matrix goes to zero in the limit $\Delta M =M_2-M_1 \rightarrow 0$. 
Indeed, from Eqs.~(\ref{condition}) and (\ref{mbb1loopA}) we find:
\begin{eqnarray}
m^{\rm 1-loop}_{\beta\beta} 
&=&\, \frac{1}{(4\, \pi\, v)^2}\,
(\theta V)_{e 1}^2\, M^3_1\,
\left\{\left [\left ( \frac{3 \log( M_1^{2}/M_{Z}^{2})}{M_1^{2}/M_{Z}^{2}-1}
+\frac{\log(M_1^{2}/M_{H}^{2})}{M_1^{2}/M_{H}^{2}-1} \right) - 
\left ( M^2_1 \rightarrow M^2_2\right )\right ] \right. 
\nonumber \\ [0.30cm]
 &-&\, \left. z(2+z)\,
\left(\frac{3 \log( M_2^{2}/M_{Z}^{2})}{M_2^{2}/M_{Z}^{2}-1}
+\frac{\log(M_2^{2}/M_{H}^{2})}{M_2^{2}/M_{H}^{2}-1} \right)
\right \}\,,
\label{mv1loopcorr1}
\end{eqnarray}
%
where $z \equiv \Delta M/M_1$, i.e., $M_2 = (1 + z) M_1$.
Note that Eq.~(\ref{mv1loopcorr1}) is valid for 
arbitrary values of $z$ and $M_1$.
In the case of $M^2_1,M^2_2 \ll M^2_{Z},M^2_{H}$ we get:
\begin{eqnarray}
m^{\rm 1-loop}_{\beta\beta} 
&=&\, \frac{(\theta V)^2_{e 1}}{(4\, \pi\, v)^2}\, M^3_1\,
\left [8\,(1+z)^2 \log (1 + z) + z(2+z)\,\left(3 \log( M_1^{2}/M_{Z}^{2})
+ \log(M_1^{2}/M_{H}^{2}) \right)
\right ]\,.\nonumber\\
\label{mv1loopcorr2}
\end{eqnarray}
%
If, in addition, $z \ll1$, this expression further simplifies to:
\begin{eqnarray}
m^{\rm 1-loop}_{\beta\beta}  
&=&\, \frac{(\theta V)^2_{e 1}}{(4\, \pi\, v)^2}\, M^3_1\, z(2+z)\,
\left [4(1+z)^2 + 3 \log( M_1^{2}/M_{Z}^{2})
+ \log(M_1^{2}/M_{H}^{2}) \right ]\,.
\label{mv1loopcorr3}
\end{eqnarray}
%
In the opposite limit, namely,  $M^2_1,M^2_2 \gg M^2_{Z},M^2_{H}$, 
$m^{\rm 1-loop}_{\beta\beta} $ takes also a rather simple form 
for $z \ll 1$. In this case,  to leading order in $z \ll 1$, 
we obtain:
\begin{eqnarray}
m^{\rm 1-loop}_{\beta\beta} 
&=&\,-\, 2\, z\,\frac{1}{(4\, \pi\, v)^2}\,
(\theta V)^2_{e 1}\, M_1\,\left (3\, M^2_{Z} + M^2_{H} \right )\,.
\label{mv1loopcorr4}
\end{eqnarray}
%

Thus, in the scheme considered here, in which 
condition (\ref{condition}) is fulfilled,
the magnitude of the one-loop correction 
 to $m^{\text{light}}_{\beta\beta}$ 
of interest, $m^{\rm 1-loop}_{\beta\beta} $,
exhibits a strong dependence on $z$.
This dependence is particularly important 
in the case when the two heavy Majorana 
neutrinos form a pseudo-Dirac pair,
$0 < \Delta M \ll M_1,M_2$, or $z\ll 1$.
In this case the ratio of the 
one-loop  correction  to 
the $0\nu\beta\beta$ decay amplitude and 
the heavy Majorana neutrino exchange 
contribution given in Eq.~(\ref{meeh2}),
$|m^{\rm 1-loop}_{\beta\beta} /m_{\beta\beta}^{\rm heavy}|$, practically
depends  only on the mass $M_1$.
As it is not difficult to show, 
for $f(A) =  0.79~(0.033)$, i.e., for $^{76}$Ge ($^{48}$Ca), 
we have $|m^{\rm 1-loop}_{\beta\beta}/m_{\beta\beta}^{\rm heavy}| \approx 1$ 
at  $M_1 \approx 15~(9.7)$ GeV.
For $M_1 > 15~(9.7)$ GeV 
($M_1 < 15~(9.7)$ GeV),  $|m^{\rm 1-loop}_{\beta\beta} |$ 
is bigger (smaller) than $|m_{\beta\beta}^{\rm heavy}|$.
\begin{figure}[t!]
\includegraphics[width=0.6\textwidth,angle=0]{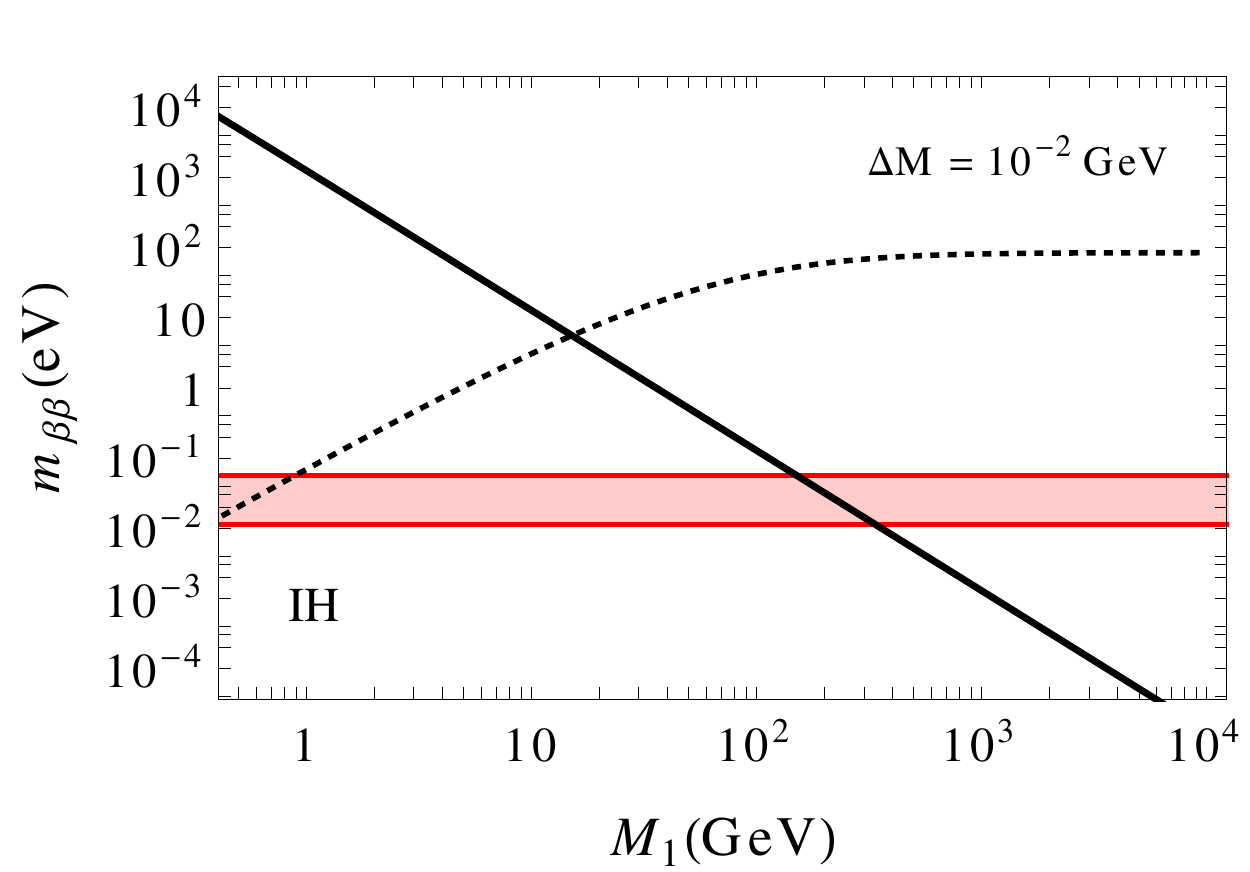} 
\caption{
\label{1loopandh}
{\small 
The contributions to 
the $0\nu\beta\beta$ decay effective Majorana mass 
due to the one-loop correction to the light neutrino mass 
matrix (dashed line) and due to the heavy Majorana neutrino 
exchange (solid line), 
$|m^{\rm 1-loop}_{\beta\beta}|$ and $|m_{\beta\beta}^{\rm heavy}|$ 
(Eqs.~(\ref{mv1loopcorr1}) and (\ref{meeh2})),
as functions of the heavy Majorana neutrino mass $M_1$,
for $\Delta M = 10^{-2}$ GeV, $|(\theta V)_{e1}|^2 = 10^{-3}$ 
and $f(A) = 0.079$ (i.e., for $^{76}$Ge).
The range of values  the effective Majorana neutrino mass can take 
in the case of light Majorana neutrino exchange and 
IH spectrum is also shown (the band in red color). 
See the text for further details.
}}
\end{figure}
%
This is illustrated in Fig.~\ref{1loopandh}, which shows the dependence 
of  $|m^{\rm 1-loop}_{\beta\beta} |$ and $|m_{\beta\beta}^{\rm heavy}|$ 
on $M_1 > 0.5$ GeV for $\Delta M = 10^{-2}$ GeV in the scheme in which 
condition~(\ref{condition}) is exactly fulfilled and 
fixing the active-sterile mixing to
the reference value of $|(\theta V)^2_{e 1}|= 10^{-3}$.
In this
plot the Higgs mass has been set to $M_H=125$ GeV.
Note, however, that given the values of $M_Z = 90$ GeV and 
$M_H = 125$ GeV, for $M_1 = 15~(9.7)$ GeV,
the factor $(4(1+z)^2 + 3 \log( M_1^{2}/M_{Z}^{2})
+ \log(M_1^{2}/M_{H}^{2}))$ in Eq. (\ref{mv1loopcorr3}) 
for  $m^{\rm 1-loop}_{\beta\beta}$
is negative. Thus, at  $M_1 = 15~(9.7)$ GeV, 
we have $m^{\rm 1-loop}_{\beta\beta}/m_{\beta\beta}^{\rm heavy} > 0$ 
(see Eq. (\ref{meeh2})), 
and therefore a cancellation, or even a partial 
compensation, between the two terms 
$m^{\rm 1-loop}_{\beta\beta} $ and 
$m_{\beta\beta}^{\rm heavy}$  in the 
$0\nu\beta\beta$ decay amplitude is impossible.

 As it should be clear from Fig.~\ref{1loopandh} and  Eqs.~(\ref{mv1loopcorr2}-\ref{mv1loopcorr4}), 
 $|m^{\rm 1-loop}_{\beta\beta}|$ grows rapidly with the 
increase of $M_1$. However, the dependence of 
$|m^{\rm 1-loop}_{\beta\beta} |$ on $z$ when 
$z << 1$ makes it possible, in principle, 
for $|m^{\rm 1-loop}_{\beta\beta} |$ to have values 
in the range of sensitivity of the current and 
next generation of  $0\nu\beta\beta$ decay experiments,
i.e., to have  $|m^{\rm 1-loop}_{\beta\beta} | \sim (0.01 - 0.10)$ eV
even for, e.g., $M_1 = 10^{3}$ GeV and 
the maximal value of $|(\theta V)^2_{e 1}| = 10^{-3}$ 
allowed by the current data.
This requires, however, exceedingly small values of 
$z$, which lead to a subleading heavy neutrino contribution.
Indeed, using the quoted values of $M_1$ and  $|(\theta V)^2_{e 1}|$, 
and taking into account that $v = 246$ GeV, 
it is not difficult to find from Eq.~(\ref{mv1loopcorr4}) that 
we can have $|m^{\rm 1-loop}_{\beta\beta} | \approx 0.01~(0.10)$ eV
for $z \approx 6\times 10^{-10}~(6\times 10^{-9})$. Such a small value of $z$ suggests a severe fine-tuning,
but it can also be understood in the context of the 
ISS scenario as a technically naturally small value of the lepton number
violating parameters of this model.

In the analyses which follow we will not assume that 
Eq.~(\ref{condition}) relating $(\theta V)_{e 1}$ and $(\theta V)_{e 2}$
is satisfied. We will use only the phenomenological constraint 
on $(\theta V)_{e 1}$ and $(\theta V)_{e 2}$  
\cite{Antusch:2006vwa,Antusch:2008tz,Atre:2009rg,Alonso:2012ji,Antusch:2014woa,Drewes:2015iva,Dinh:2012bp,Cely:2012bz}. Notice,
however, that for values of the Casas-Ibarra parameter $|\gamma| \gtrsim 6$ (see Eqs. (\ref{thetaR}), 
(\ref{RissNH}) and (\ref{RissIH})), the relation given in Eq.~(\ref{condition}) is effectively satisfied.

\subsection{One-loop Generalisation of the Casas-Ibarra Parametrization}
\label{CIloop}

In order to make sure that we generate the correct light neutrino 
mixing pattern, it is useful to generalise the
 Casas-Ibarra parametrization introduced in the previous section 
including the one-loop correction to the neutrino mass matrix. 
Taking into account the expression (\ref{mv1loopcorr0}) for 
$(m^{\rm 1-loop}_{\nu})_{\ell\ell'}$ 
in the basis in which the RH neutrino mass is diagonal,  
Eq.~(\ref{mvtot}) takes the explicit form: 
\begin{eqnarray}
	(m_{\nu})_{\ell\ell'}&=&-\,(m_{D}\,V)_{\ell k} \left[  M^{-1}_k 
-\frac{1}{(4\, \pi\, v)^2}\,M_k\,
	\left(\frac{3 \log( M_k^{2}/M_{Z}^{2})}{M_k^{2}/M_{Z}^{2}-1}
+\frac{\log( M_k^{2}/M_{H}^{2})}{ M_k^{2}/M_{H}^{2}-1} \right)\right]\,
(V^{T}\,m_{D}^{ T})_{k\ell'}\,
\nonumber\\
	&\equiv& 
-\,(m_{D}\,V)_{\ell k}\,\Delta_k^{-1}\,(V^{T}\,m_{D}^{T})_{k \ell'}\;
=\;(U_{\rm PMNS}^*\,\text{diag}(m_1,m_2,m_3)\,U_{\rm PMNS}^\dagger)_{\ell \ell'}\,.
\label{mv1loopcorr}
\end{eqnarray}
%
Hence, in analogy to the tree-level contribution, 
we have now
\be
\left(\pm i \,\hat m^{-1/2} \,U_{\text{PMNS}}^\dagger \,\theta V \,\hat{M}\,\Delta^{-1/2}\right)\,\left(\pm i \,\hat m^{-1/2}\, U_{\text{PMNS}}^\dagger \,\theta V\,\hat{M}\, \Delta^{-1/2}\right)^T
\equiv R\, R^T=1\,.\label{CIloop1}
\ee
%
Thus, we get the following expression for the 
heavy Majorana neutrino couplings in the weak charged current, or 
equivalently, for the  active-sterile 
neutrino mixing, at one-loop order:
\be
\theta V = \mp i\,  U_{\text{PMNS}}\, \hat m^{1/2}\, R\, \Delta^{1/2}\,\hat{M}^{-1}\,.
\label{thetaR2}
\ee
%
In the numerical analysis reported in section~\ref{BetaBeta} we will make use 
of this parametrization of  $\theta V$, with $R$ given 
in (\ref{RNO}) and (\ref{RIO}), in order to include the one-loop corrections
to the light neutrino masses and at the same time ensure that all 
the neutrino mixing parameters match with their experimental values.

\begin{figure}[t!]
\begin{tabular}{cc}
\includegraphics[width=0.5\textwidth,angle=0]{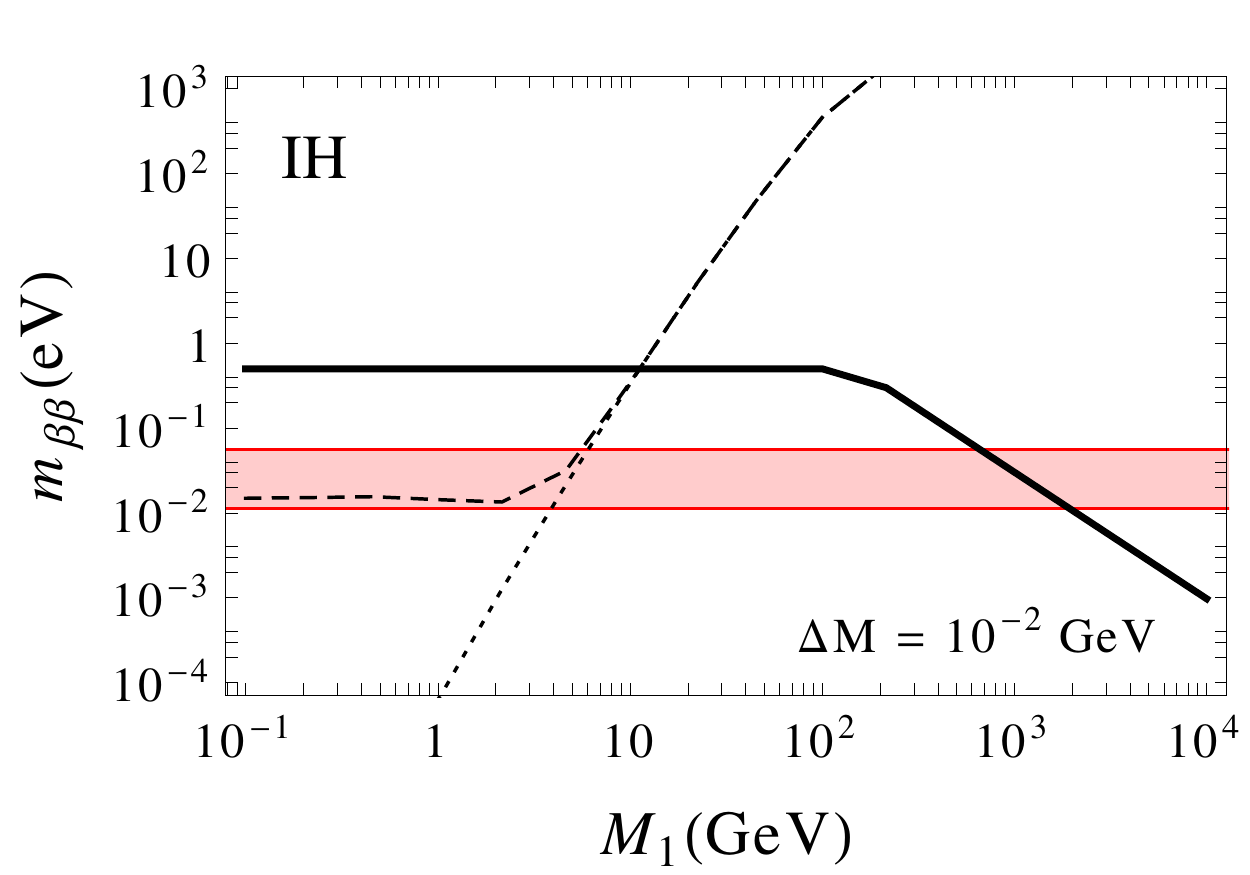} &
\includegraphics[width=0.5\textwidth,angle=0]{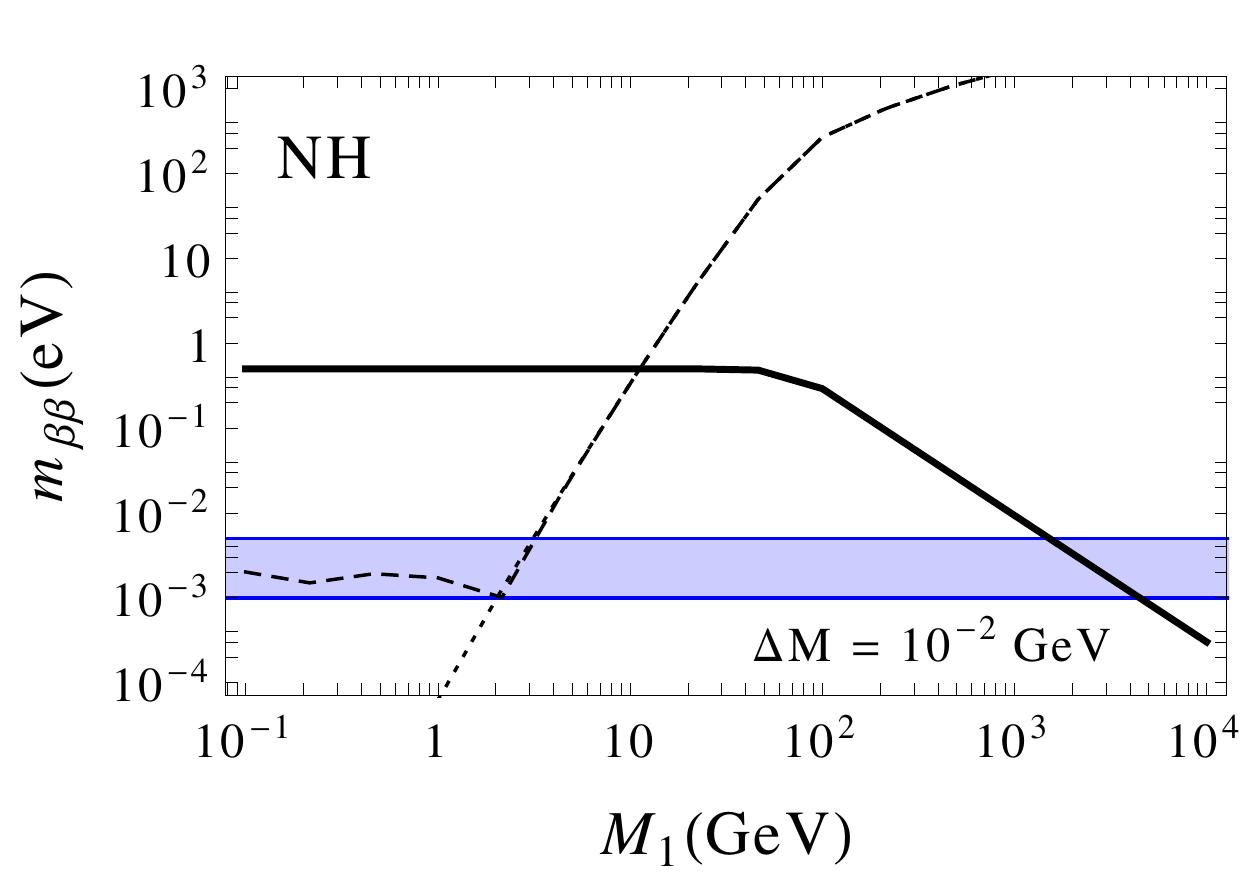} 
\end{tabular}
\caption{
\label{mbbMax}
{\small 
Maximum value of the contribution to the $0\nu\beta\beta$ decay effective Majorana mass due to the heavy 
Majorana neutrino exchange $|m_{\beta\beta}^{\rm heavy}|$ (solid thick line) for $^{76}$Ge and $\Delta M =10^{-2}$ GeV in the 
IH (left panel) and NH (right panel) case, including the following constraints: $|m_{\beta\beta}^{\rm heavy}|\leq 0.5$ eV 
and $|(\theta V)_{e 1}|^2 + |(\theta V)_{e 2}|^2  \leq 2\times10^{-3}$. The corresponding values of the contributions to the $0\nu\beta\beta$ decay 
effective Majorana mass due to the tree-level (dashed line) and one-loop correction (dotted line)
to the light neutrino mass matrix, $|m^{\rm tree}_{\beta\beta}|$ and $|m^{\rm 1-loop}_{\beta\beta}|$, are also shown. The range of 
values  the effective Majorana mass can take in the case of light Majorana neutrino exchange and IH (NH) spectrum is shown in the red (blue) band. See the text for further details.
}}
\end{figure}

In Fig.~\ref{mbbMax} we illustrate the interplay between the contributions to 
the $0\nu\beta\beta$ decay effective Majorana neutrino mass due to the heavy Majorana neutrino 
exchange, $|m_{\beta\beta}^{\rm heavy}|$, the  tree-level light neutrino masses, $|m^{\rm tree}_{\beta\beta}|=|(m^{\rm tree}_{\nu})^*_{ee}|$, and the one-loop correction to 
the light neutrino mass matrix, $|m^{\rm 1-loop}_{\beta\beta}|=|(m^{\rm 1-loop}_{\nu})^*_{ee}|$, using the generalised Casas-Ibarra parametrization derived
above. In particular, we have maximised $|m_{\beta\beta}^{\rm heavy}|$ over the free parameters of the model ($\theta_{45}$, $\gamma$ and the Dirac and Majorana
phases of the PMNS matrix), in order to show the maximum heavy neutrino contribution to the process (solid thick line)
as a function of $M_1$ for $\Delta M =10^{-2}$ GeV and fixing the already measured PMNS parameters and neutrino squared mass differences to the best fit values 
given in \cite{nufit}. The Higgs mass has been set to $M_H=125$ GeV. In the plot we show the corresponding value of the separate contributions associated to the tree-level (dashed line) and one-loop correction  (dotted line) to the light neutrino mass matrix. We also impose the following constraints: $|m_{\beta\beta}^{\rm heavy}|\leq 0.5$ eV and $|(\theta V)_{e 1}|^2 + |(\theta V)_{e 2}|^2\leq 2\times10^{-3}$. 

From Fig.~\ref{mbbMax} we conclude that for $M_1\lesssim 1$ GeV the one-loop correction is subleading for $\Delta M =10^{-2}$ GeV, being the tree-level contribution 
the one responsible for the light neutrino mass generation. At the same time, in that region the heavy neutrino contribution to the $0\nu\beta\beta$ decay effective Majorana neutrino mass 
can be sizable and larger than the one from light neutrino exchange. According to the estimate given in Fig.~\ref{estimate}, for $M_1\lesssim 1$ GeV there is no
need of any enhancement of the active-sterile mixing with respect to the naive seesaw scaling in order to obtain a sizable $|m_{\beta\beta}^{\rm heavy}|$. 
However, around $M_1\sim 2$ GeV, the one-loop correction starts to be of the same size as the value of the light neutrino contribution dictated by neutrino oscillation
data. Indeed, this correction increases with $M_1$ in such a way that in order to stabilise the light neutrino mass and mixing, a fine-tuned cancellation 
between the tree-level and one-loop correction is required. This is reflected in the fact that for $M_1\gtrsim 5$ GeV the dotted and dashed lines merge. 
Therefore, as it is shown in Fig.~\ref{mbbMax}, for $5$ GeV $\lesssim M_1\lesssim 1$ TeV a sizable $|m_{\beta\beta}^{\rm heavy}|$ can in principle be  realised, but a fine-tuned cancellation between the tree-level and one-loop 
contributions to the light neutrino masses is also necessary. 

Note that the bound $|m_{\beta\beta}^{\rm heavy}|\leq 0.5$ eV imposed by us can be saturated for $M_1\lesssim 100$ GeV. At 
$M_1 = 10$ GeV, for instance, we have $|m_{\beta\beta}^{\rm heavy}| = 0.5$ eV for $|(\theta V)_{e1}|^2 + |(\theta V)_{e2}|^2 \simeq 0.8\times 10^{-4}$,
where we have used $f(A) = 0.079$ corresponding to $^{76}$Ge. For $M_1 \gtrsim 100$ GeV the 
maximum value of $|m_{\beta\beta}^{\rm heavy}|$ decreases with $M_1$ since an active-sterile mixing $|(\theta V)_{ei}|^2$ bigger than $2\times10^{-3}$ 
would be required in order to saturate the bound. 

It is interesting that the solid line and the blue and red bands 
in Fig.~\ref{mbbMax} intersect around $M_1\sim 10^3$ GeV. 
This implies that in the case of NH neutrino mass spectrum, 
the effective Majorana neutrino mass $|m_{\beta\beta}|$  can be larger at 
$0.1~{\rm GeV}\lesssim  M_1\lesssim 10^3$ GeV
than that predicted in the case of the light neutrino exchange 
mechanism. In particular, it can be in the range of sensitivity 
of the experiments aiming to probe the range of 
values of the effective Majorana mass corresponding to the 
IH and quasi-degenerate (QD) light neutrino mass spectra 
(see, e.g., \cite{PDG2014}). 
In the case of the IH light neutrino mass spectrum, 
the indicated result implies that at 
$M_1\lesssim 10^3$ GeV there can be, in principle, 
a significant interplay between the 
light and heavy Majorana neutrino exchange contributions in 
the effective Majorana mass, as discussed in detail in 
\cite{Ibarra:2011xn} and summarised by us  
at the end of subsection II.A 
(see the paragraph before the last in subsection II.A). 
More specifically, due to this interplay
of the light and heavy Majorana neutrino contributions,
$|m_{\beta\beta}|$ can be larger (smaller)
than that predicted in the case of the exchange of light neutrinos
with IH mass spectrum and 
$|m_{\beta\beta}|$ will exhibit a dependence on the 
atomic number $A$ of the decaying nucleus.
It should be mentioned that, 
given the already high level of fine-tuning 
required for the cancellation between the tree-level and one-loop light 
neutrino contributions in $m_{\beta\beta}$, 
an additional cancellation between the light and heavy 
Majorana neutrino contributions would suggest further 
fine-tuning.

The main features of Fig.~\ref{mbbMax}  also  appear
for larger splittings $\Delta M$. 
In particular, the necessity of fine-tuned cancellation 
between the tree-level and one-loop 
correction to the light neutrino mass matrix is present also 
in this case. 
The level of the fine-tuning required increases 
with $M_1$, as we will show in Section \ref{BetaBeta}.

\subsection{Radiative Corrections to the ESS and ISS Scenarios}

In this section, we compute the one-loop contribution to the 
effective Majorana neutrino mass in the ESS and ISS limits of 
the seesaw Lagrangian  (\ref{typeI}) with two RH neutrinos. 
Accordingly, we apply the parametrization of the Dirac and Majorana 
mass matrices reported in Eq.~(\ref{Mnu2}) to  the general 
expression given in Eq.~(\ref{mv1loopB}).
The exact result of the one-loop contribution in terms of 
the parameters introduced in (\ref{Mnu2}) is reported in Appendix~\ref{App}.

For the ESS scenario we have at leading order in $\Lambda/\mu'$
\bea
m^{\rm 1-loop}_{\beta\beta}&\approx&\frac{\mu'}{2}\frac{y_{1e}^2}{\left(4\,\pi\right)^2}
\left(\frac{3\ln\left(\mu'^2/M_Z^2\right)}{\mu'^2/M^2_Z-1}+\frac{\ln\left(\mu'^2/M^2_H\right)}{\mu'^2/M^2_H-1}
\right)\,.
\label{meff1loopESS}
\eea
%
Notice that for $\mu'\gg M_H, M_Z$, this expression 
reduces to
\be m^{\rm 1-loop}_{\beta\beta}\;\approx\;\frac{y_{1e}^2}{\left(4\,\pi\right)^2}
\left(\frac{3\,M_Z^2}{2\,\mu'}\ln\left(\mu'^2/M_Z^2\right)+\frac{M^2_H}{2\,\mu'}\ln\left(\mu'^2/M^2_H\right)\right)\,.
\ee
%
Therefore, when $\mu'\gg M_H, M_Z$, since the lepton number violating scale 
$\mu'$ is introduced at high energies, the one-loop contribution 
to the light neutrino masses appears to be suppressed as $1/\mu'$, as expected.

In the ISS realisation, i.e. for $\epsilon\,v,\,\mu,\,\mu'\ll \Lambda$, 
we obtain
\bea
 m^{\rm 1-loop}_{\beta\beta}&\approx&\frac{1}{\left(4\,\pi\right)^2}\left(\epsilon \,\Lambda\,y_{1e}\,y_{2e}\,-\,\frac{\mu}{2}\,y_{1e}^2\right)
\left(\frac{3\ln\left(\Lambda^2/M_Z^2\right)}{\Lambda^2/M^2_Z-1}+\frac{\ln\left(\Lambda/M^2_H\right)}{\Lambda^2/M^2_H-1}\right)
\label{meff1loopISS}\\
&-&\frac{\mu+\mu'}{2}\frac{y_{1e}^2}{(4\,\pi)^2}
\left(\frac{4M_H^2M_Z^2-\Lambda^2\left(M_H^2+3M_Z^2\right)}{\left(\Lambda^2-M^2_Z\right)\left(\Lambda^2-M^2_H\right)}
+\frac{\ln\left(\Lambda^2/M_H^2\right)}{\left(\Lambda^2/M_H^2-1\right)^2}
+\frac{3\ln\left(\Lambda^2/M_Z^2\right)}{\left(\Lambda^2/M_Z^2-1\right)^2}\right)\,.
\nonumber
\eea

It is remarkable that in the ESS limit with $\mu'\lesssim  M_H, M_Z$ and in the ISS limit the one-loop correction 
to the light neutrino masses has a contribution proportional to $\mu'$. This dependence on $\mu'$ is very relevant since at one-loop 
the light neutrino contribution to the $0\nu\beta\beta$ decay amplitude does depend directly on $\mu'$, as for the heavy contribution in (\ref{mbbESS2})
and (\ref{mbbISS}). This makes much more difficult to obtain a dominant contribution from the RH neutrinos in this limit, unless a fine-tuning of 
the seesaw parameters is introduced to guarantee the smallness of the neutrino masses as it was indeed already shown in Fig. \ref{mbbMax}.

\mathversion{bold}
\section{Large heavy neutrino contribution to $0\nu\beta\beta$ decay}
\mathversion{normal}\label{BetaBeta}

In this section, we will  address in more detail the question if  
the RH neutrinos can eventually give a sizable contribution to
the $0\nu\beta\beta$ decay rate. As we have already mentioned, 
cosmological constraints close the mass window of 
$M<100$ MeV~\cite{Hernandez:2013lza,Hernandez:2014fha} 
and thus only if the RH neutrino masses are larger 
than $100$ MeV, a direct contribution to the process of interest 
can be expected. 

Following the notation in Ref.~\cite{Blennow:2010th}, 
the $0\nu\beta\beta$ decay rate can be written as 
\begin{equation}
\frac{\Gamma_{0\nu\beta\beta}}{\ln2}=
G_{01}\left|\sum_{j}U_{ej}^{2}\frac{m_j}{m_e}\mathcal{M}^{0\nu\beta\beta}(m_j)\right|^{2},
\label{decayrate}
\end{equation}
%
where $G_{01}$ is a well-known kinematic factor, $U$ is the unitary 
matrix given in Eq. (\ref{U}) which diagonalizes the 
complete neutrino mass matrix, 
$m_{j}$ are the corresponding eigenvalues, i.e., 
the neutrino masses (light and heavy), and 
$\mathcal{M}^{0\nu\beta\beta}$ are the Nuclear Matrix 
Elements (NMEs) associated with the process. Notice that 
the NMEs depend on the mass of the neutrino mediating the 
process since the dependence on the 
neutrino propagator is already included in the NMEs computation. 
The sum should be made over all the neutrino masses, 
including the heavy ones. In the following 
we will use the NMEs data provided in \cite{Blennow:2010th}. 
In particular, we will consider the NMEs computed for the $^{76}$Ge. 
However, we have checked that the conclusions of our 
analysis do not significantly change considering a different nucleus.

We will use the modified Casas-Ibarra parametrization of the
active-sterile neutrino mixing given in Eq.~(\ref{thetaR2}),
to compute the full effective Majorana neutrino mass 
$m_{\beta\beta}$, which is given by the sum of  the 
contributions from the exchange of the light and heavy Majorana neutrinos.
In this way, we include in the computation the effect of the
one-loop correction to the light neutrino masses, reproducing 
at the same time the correct neutrino oscillation parameters. 
We will also take into account the relevant bounds on the active-sterile 
mixing which come from direct searches, charged lepton flavour violation
and non-unitarity constraints
~\cite{Antusch:2006vwa,Antusch:2008tz,Atre:2009rg,Alonso:2012ji,Antusch:2014woa,Drewes:2015iva,Dinh:2012bp,Cely:2012bz}.
Notice that the inclusion of such bounds  guarantees the perturbativity of the neutrino
Yukawa couplings for any value of RH neutrino masses considered in this paper.

\begin{figure}[t!]
\begin{tabular}{ccc}
  \includegraphics[width=0.33\textwidth,angle=0]{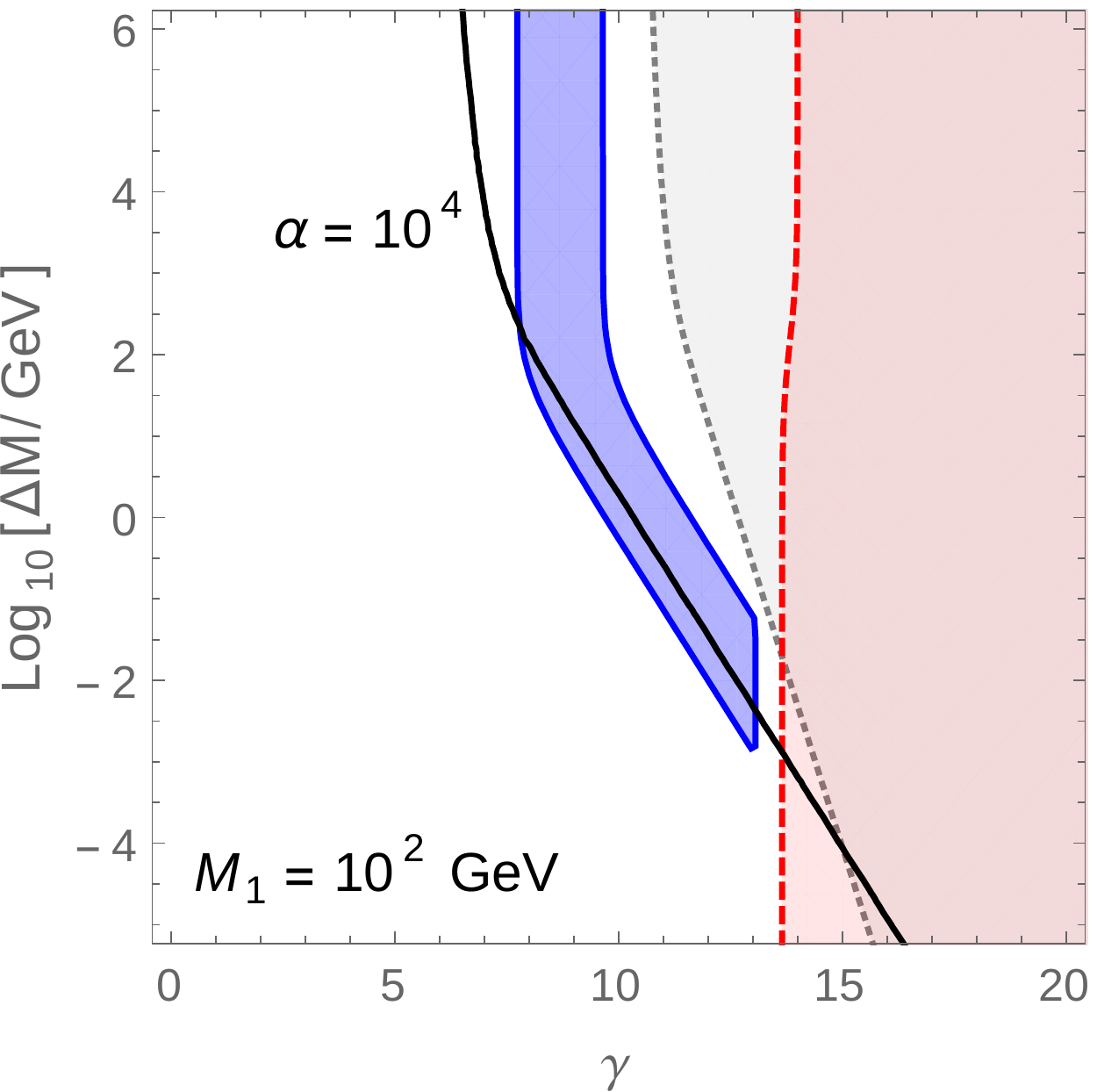} &
    \includegraphics[width=0.33\textwidth,angle=0]{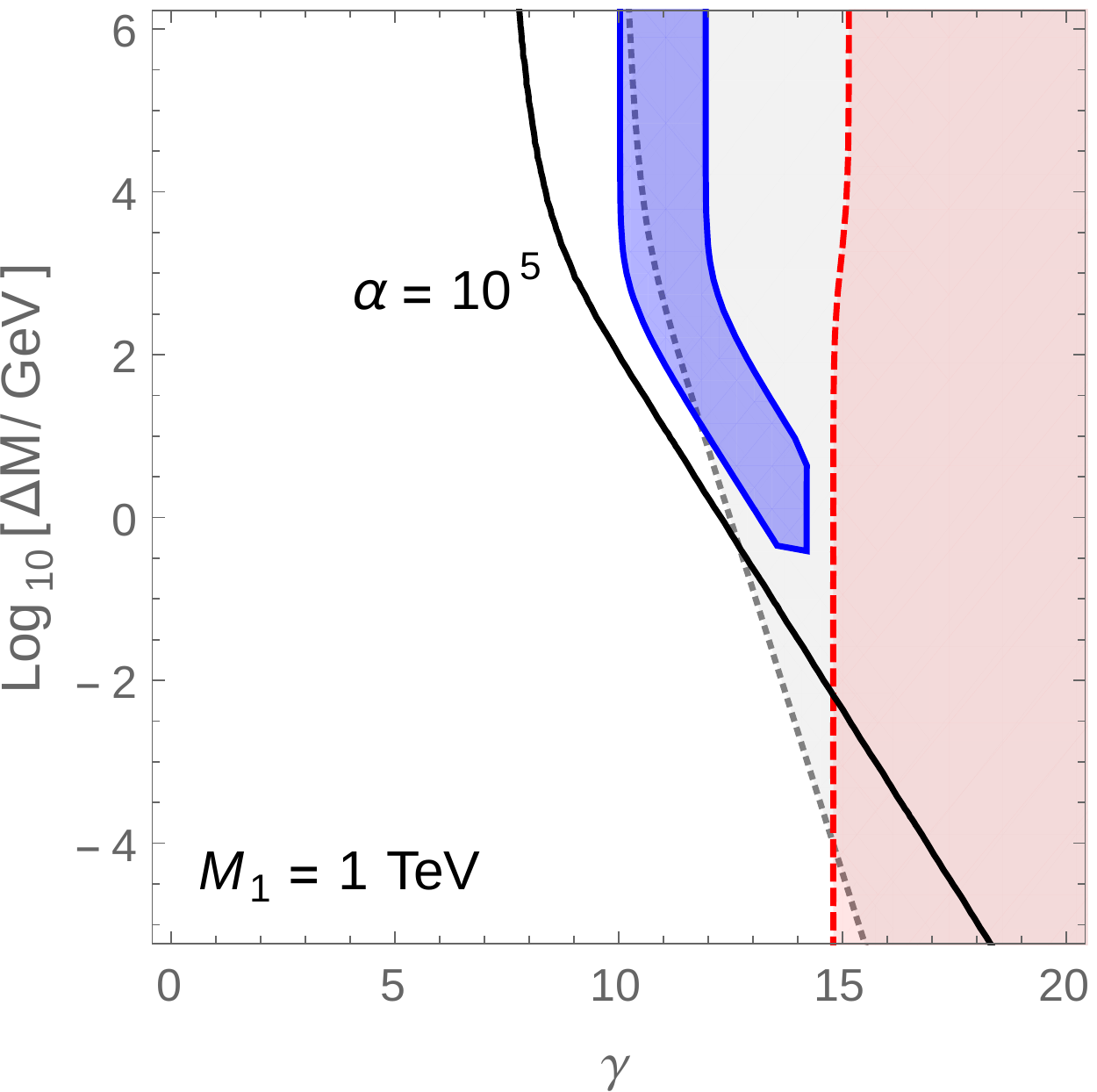}&
\includegraphics[width=0.33\textwidth,angle=0]{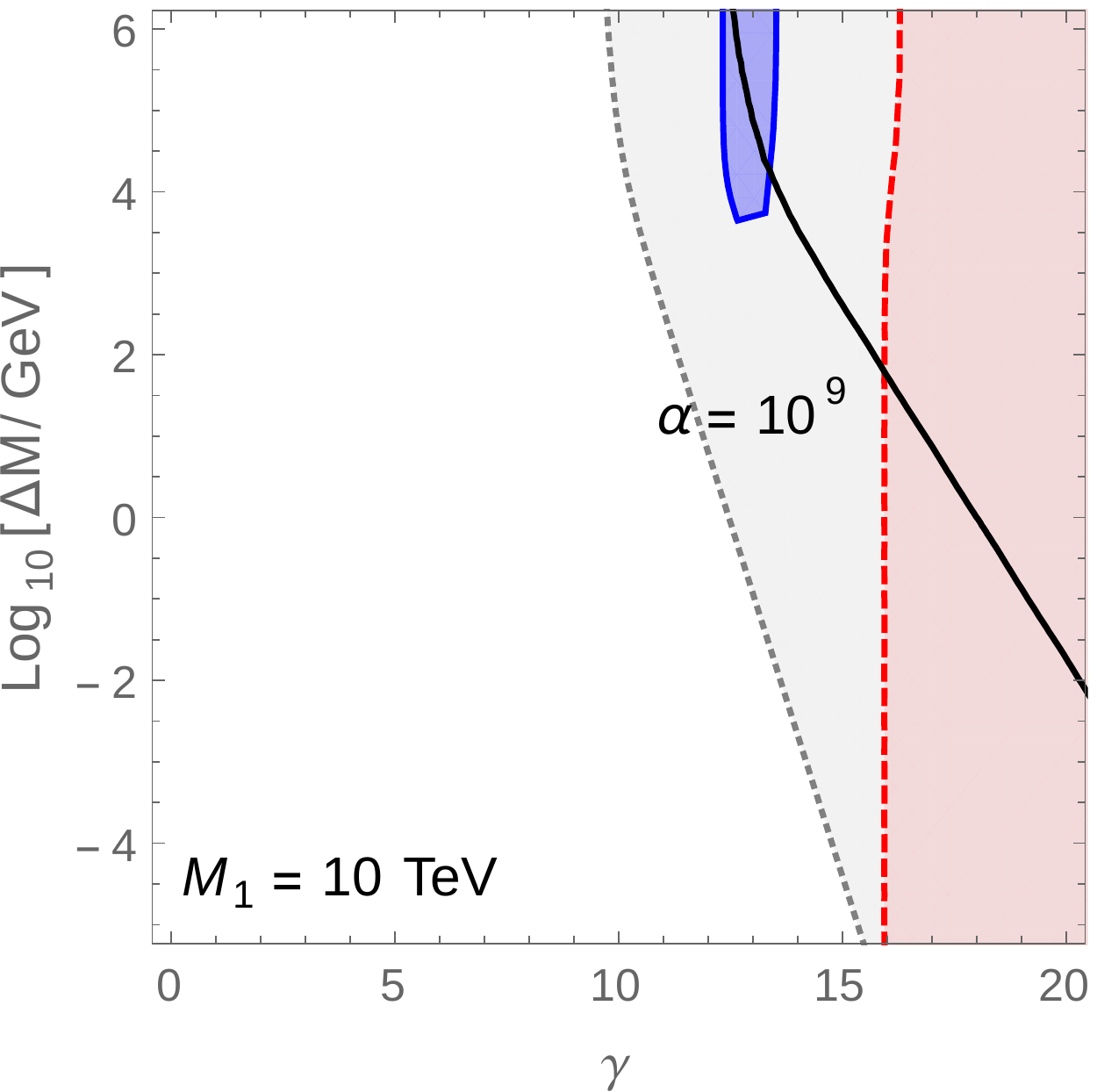}
 \end{tabular}
\begin{tabular}{ccc}
  \includegraphics[width=0.33\textwidth,angle=0]{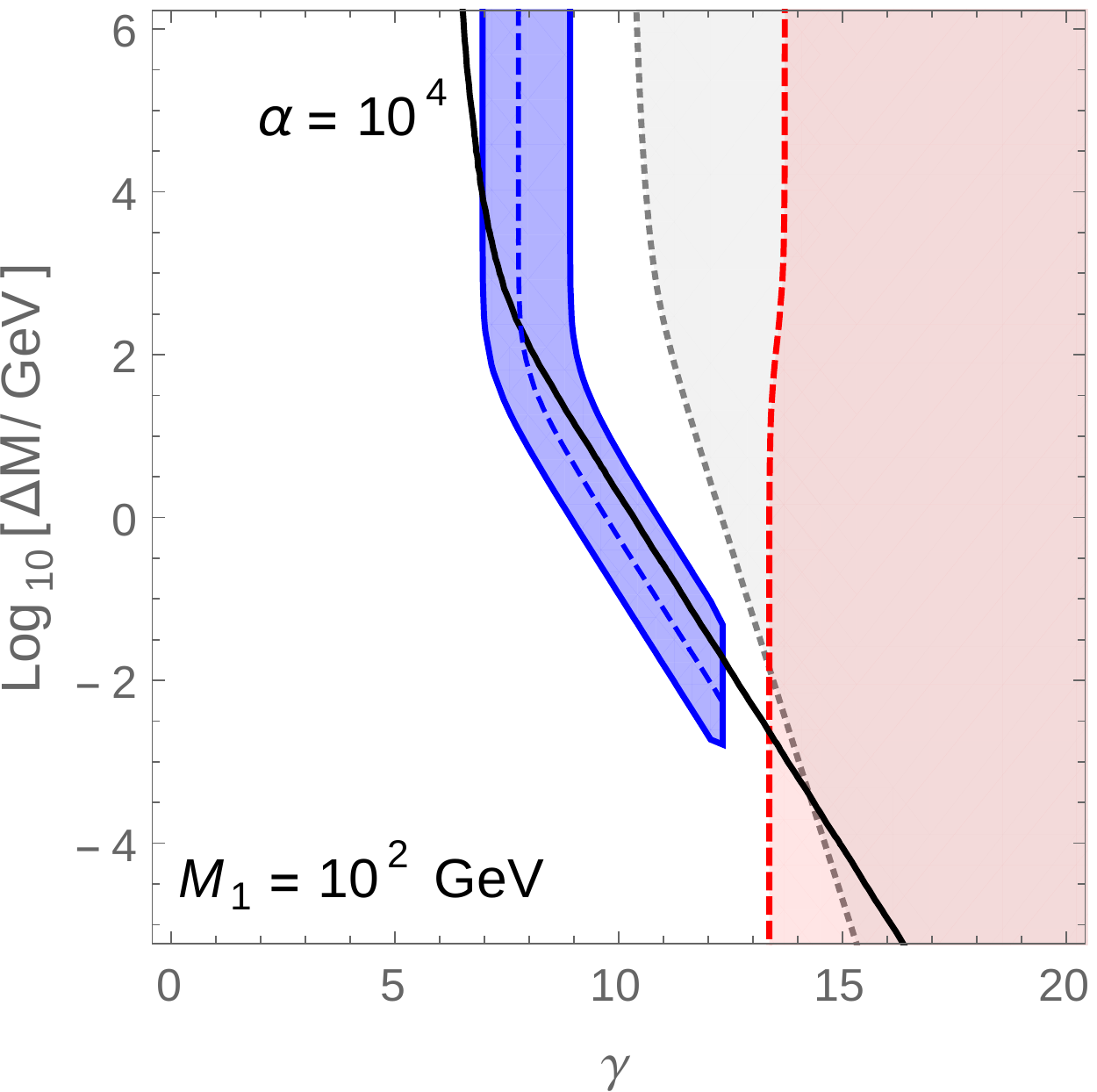} &
    \includegraphics[width=0.33\textwidth,angle=0]{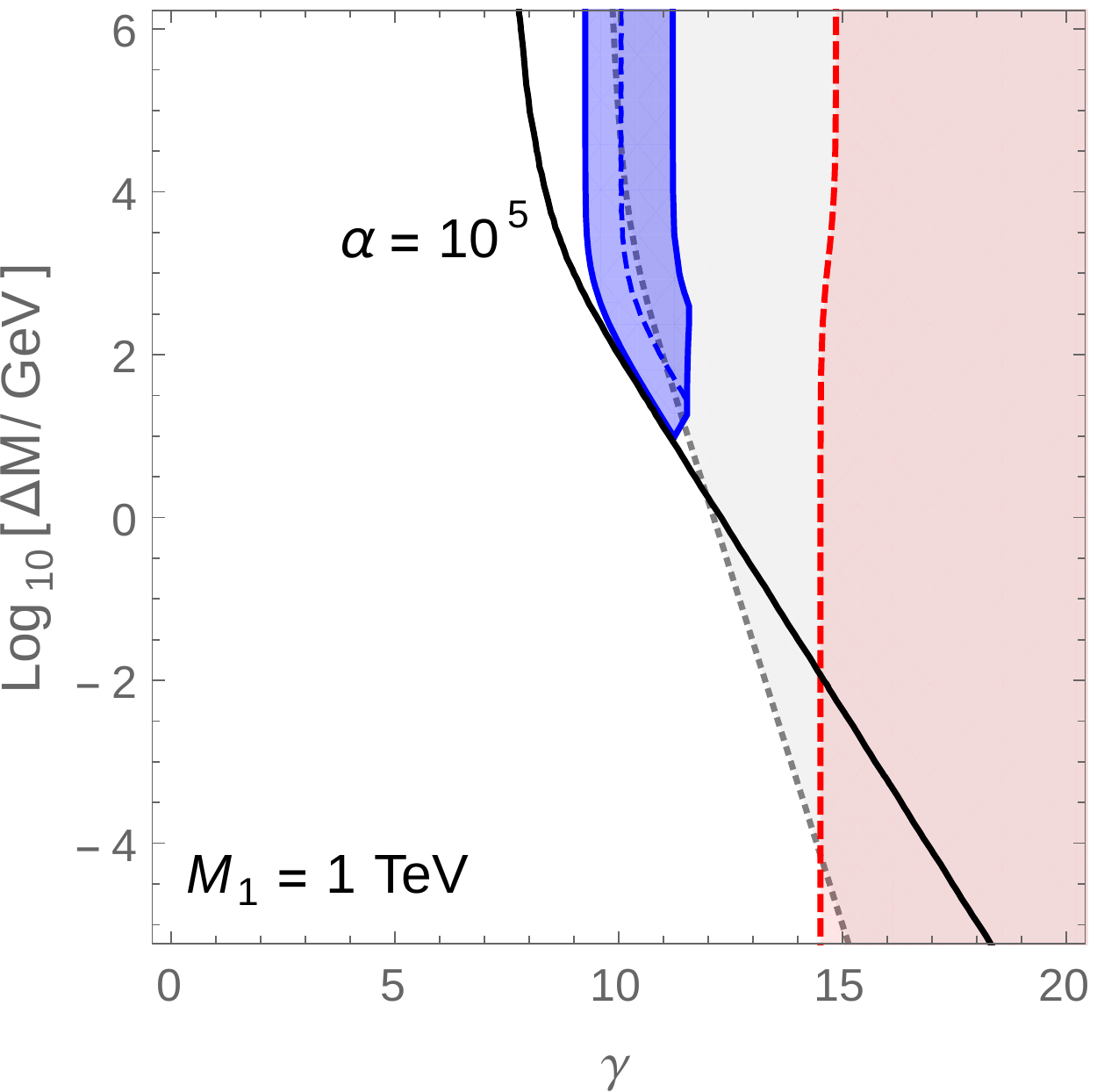}&
\includegraphics[width=0.33\textwidth,angle=0]{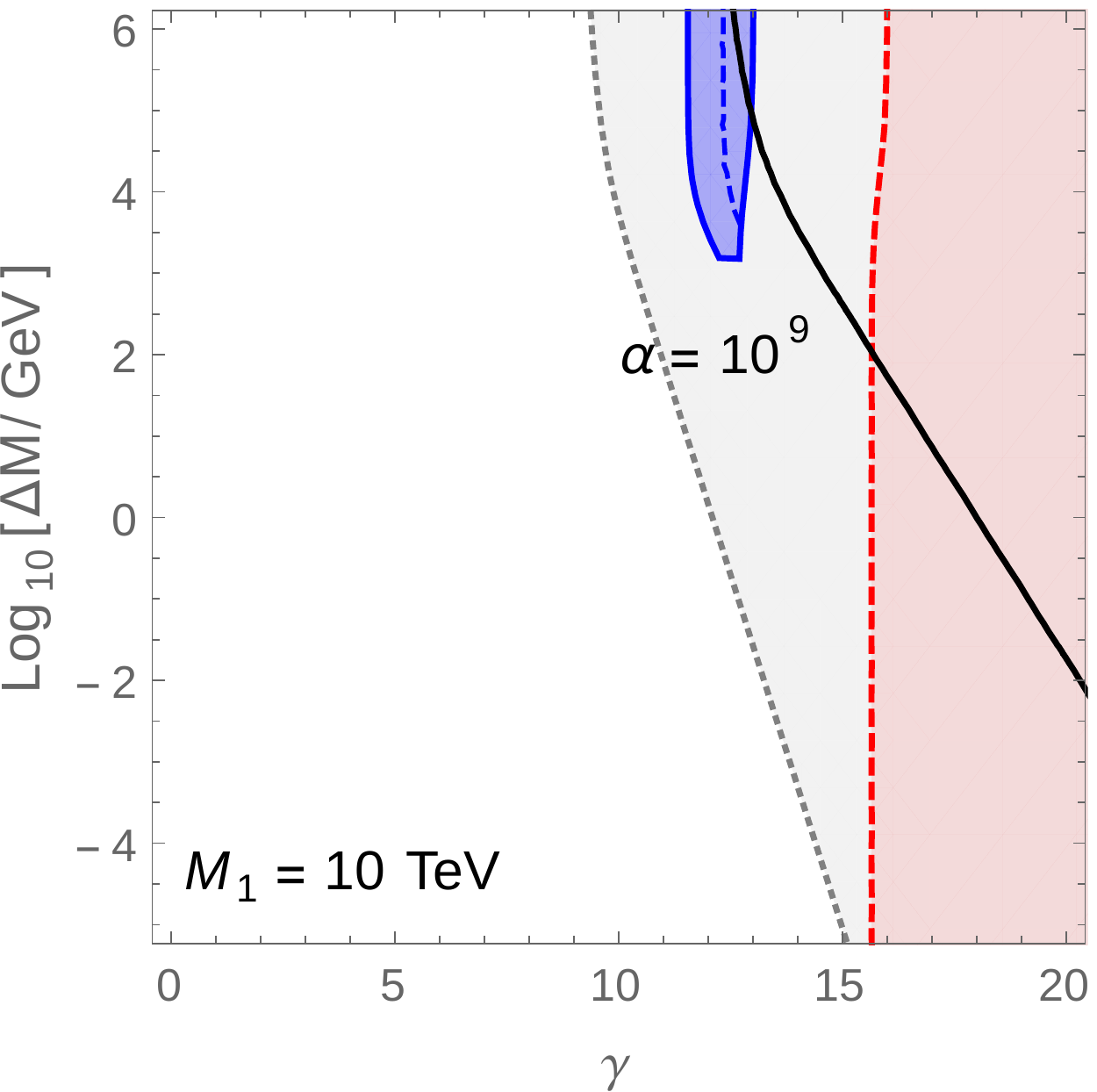}
 \end{tabular}
\caption{\label{fig2}
{\small \mathversion{bold}
\textbf{Neutrinoless double beta decay ($M_1\geq 100~\text{GeV}$).}
\mathversion{normal} The blue shaded areas in 
the top panels (down panels) 
represent the region of the parameter space in which we have 
$10^{-2}~\text{eV}<|m_{\beta\beta}^{\rm light}\,+\,m_{\beta\beta}^{\rm heavy}|<0.5~\text{eV}$ 
($10^{-2}~\text{eV}<|m_{\beta\beta}^{\rm heavy}|<0.5~\text{eV}$) 
 in the case of NH (IH) neutrino mass spectrum 
with  the active-sterile mixing (or couplings) $(\theta V)_{\ell k}$ 
satisfying the bounds form direct searches, charged lepton flavour violation  and non-unitarity constraints. 
The black solid line stands for different values of the parameter 
$\alpha \equiv |m^{\rm 1-loop}_{\beta\beta}|/|m_{\beta\beta}^{\rm light}|$, 
which quantifies the fine-tuning required in order to achieve 
the cancellation between the one-loop and tree-level 
contributions to the light neutrino masses.
In the region to the right of the red dashed line 
the ratio between the leading order and the next to 
leading order contributions to the light
neutrino masses in the seesaw expansion is smaller than $10$. 
The gray region to the right of the dotted line corresponds to 
$y_{1e}^2\, m^{\rm 1-loop}_{\beta\beta}> 
16\,\pi^2\,m_{\beta\beta}^{\rm light}$. The blue dashed line corresponds 
to $|m_{\beta\beta}^{\rm heavy}|=0.05$ eV.
The measured neutrino oscillation parameters are fixed to the central 
values reported in~\cite{nufit}.}}
\end{figure}

\begin{figure}[t!]
\begin{tabular}{ccc}
  \includegraphics[width=0.33\textwidth,angle=0]{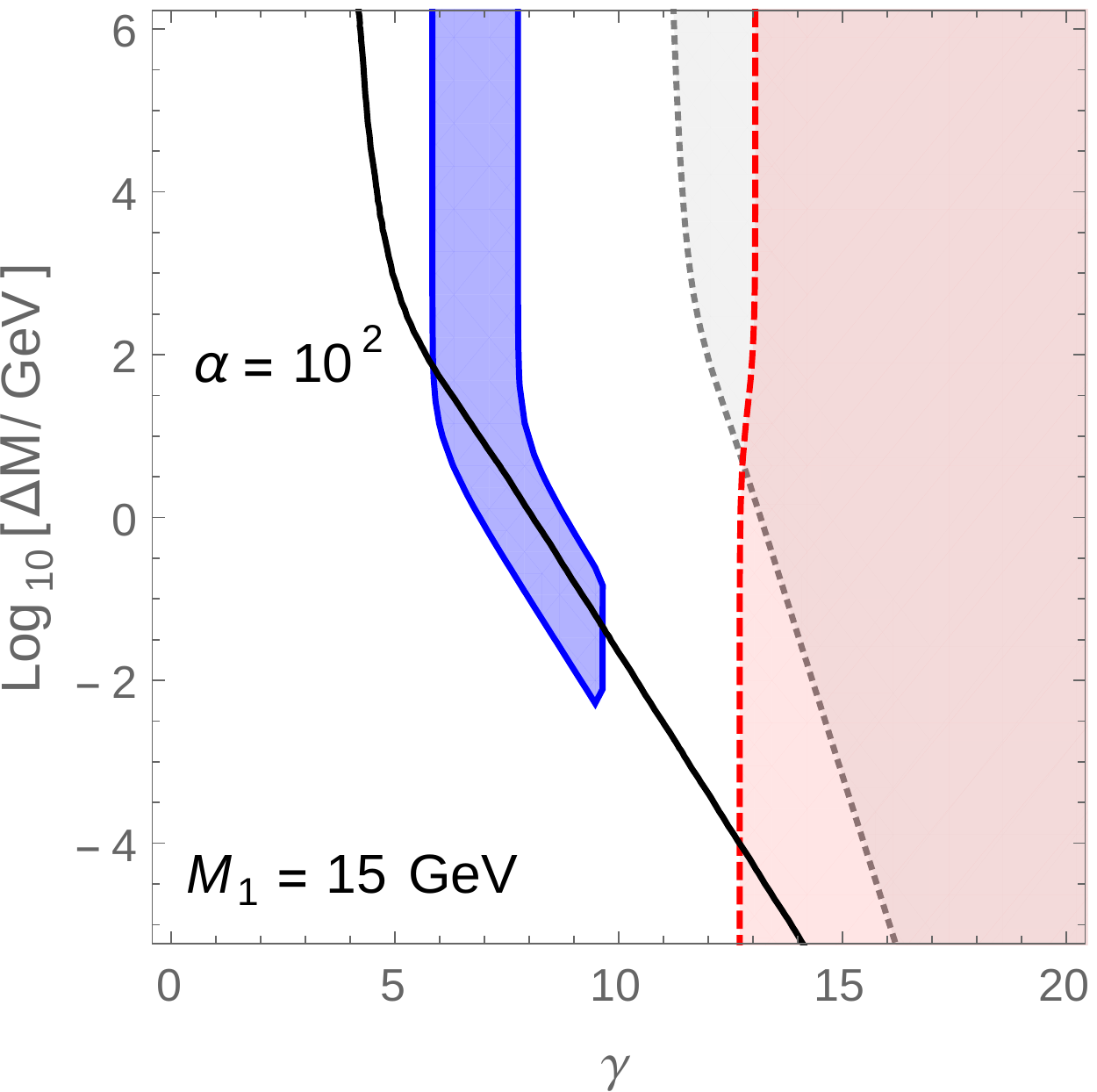} &
    \includegraphics[width=0.33\textwidth,angle=0]{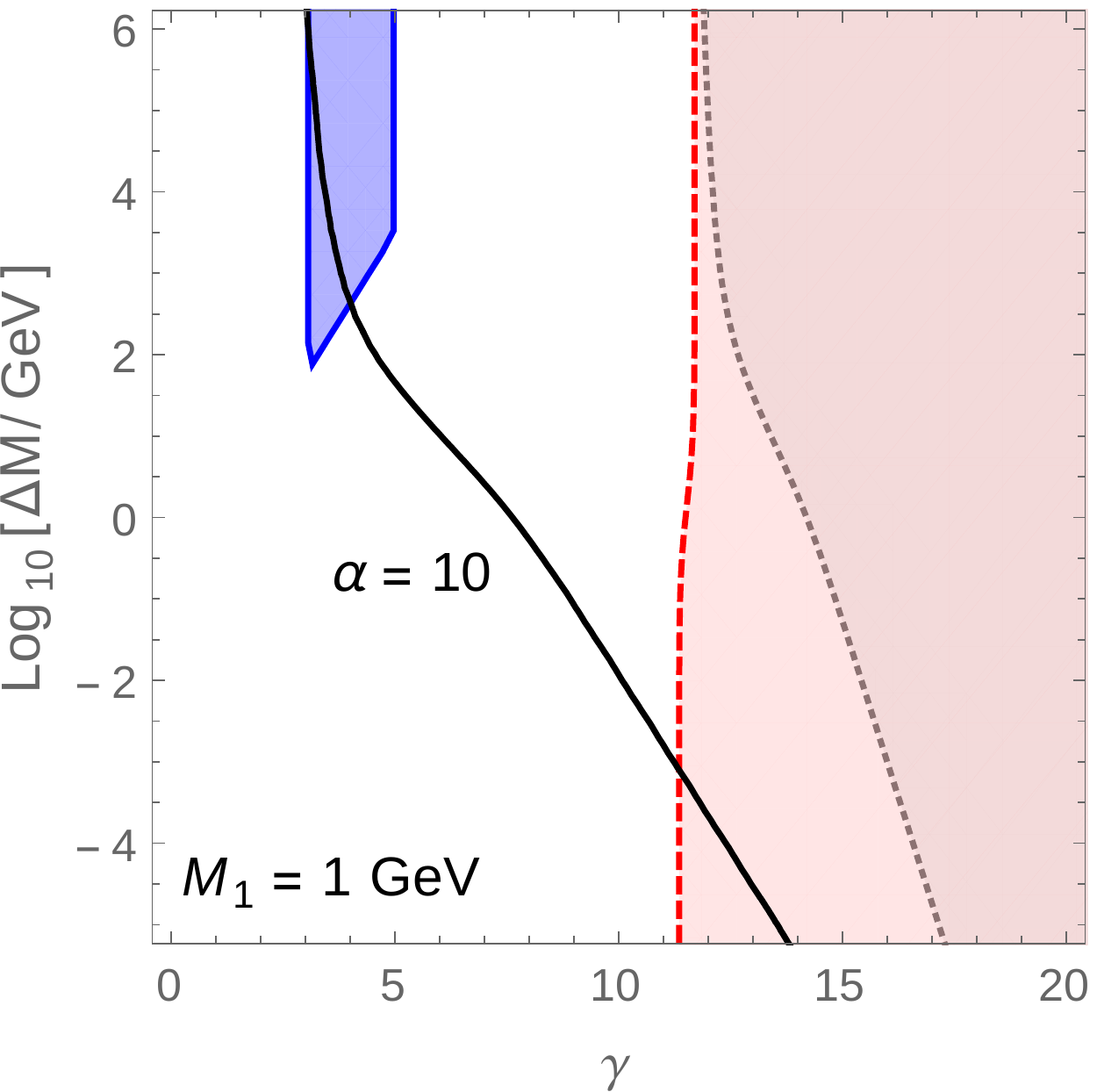}&
\includegraphics[width=0.33\textwidth,angle=0]{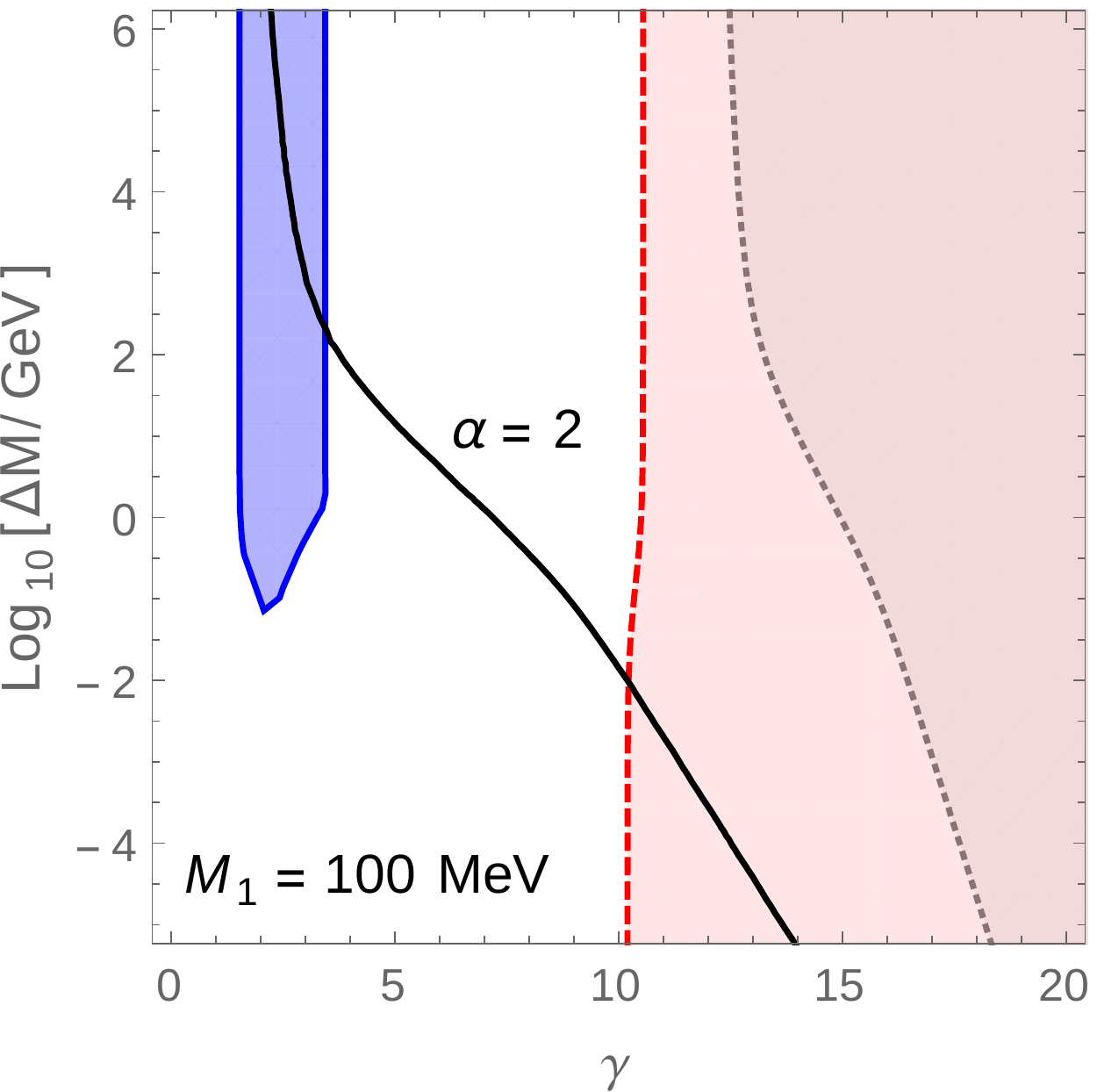}
 \end{tabular}
\begin{tabular}{ccc}
  \includegraphics[width=0.33\textwidth,angle=0]{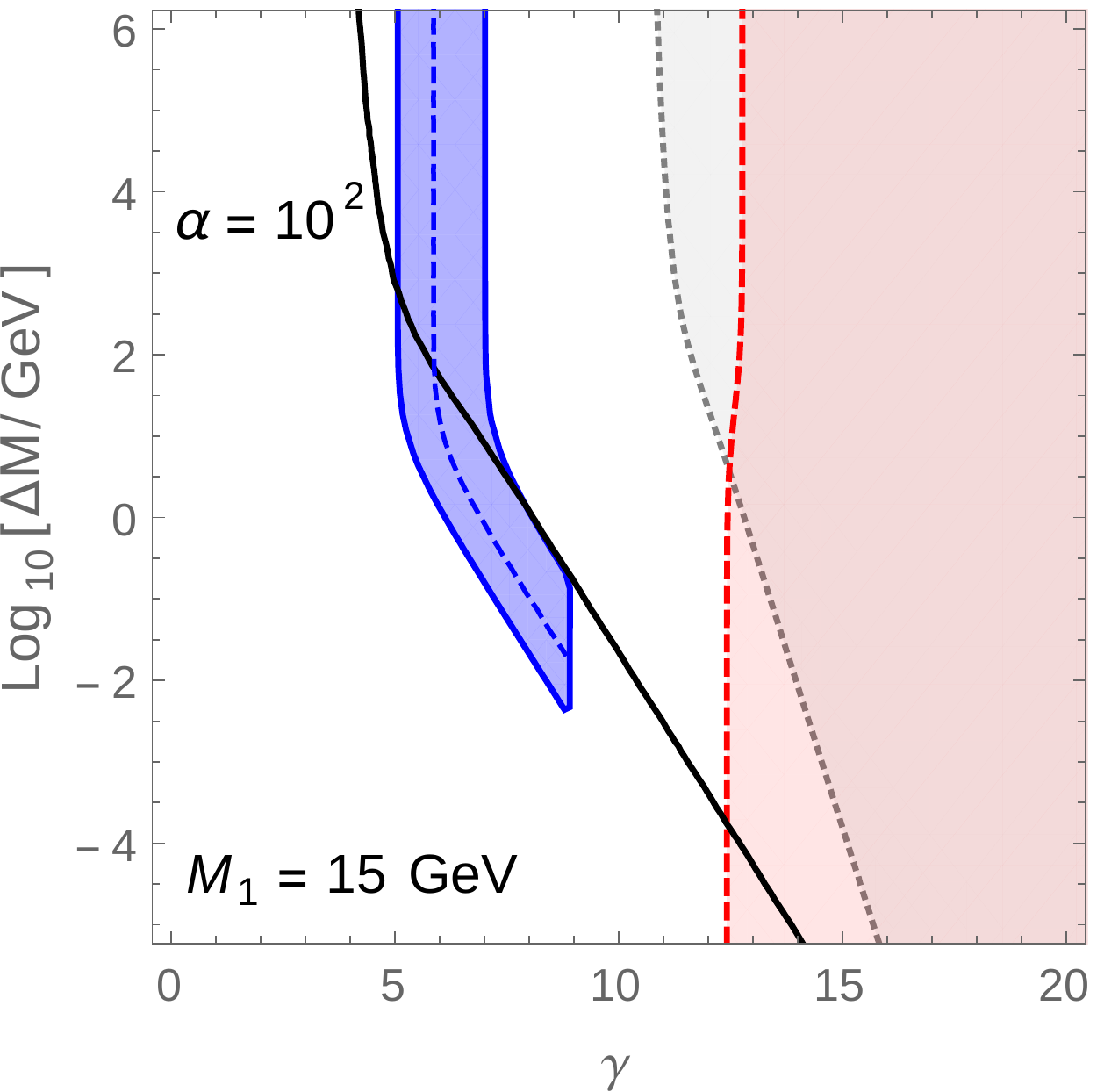} &
    \includegraphics[width=0.33\textwidth,angle=0]{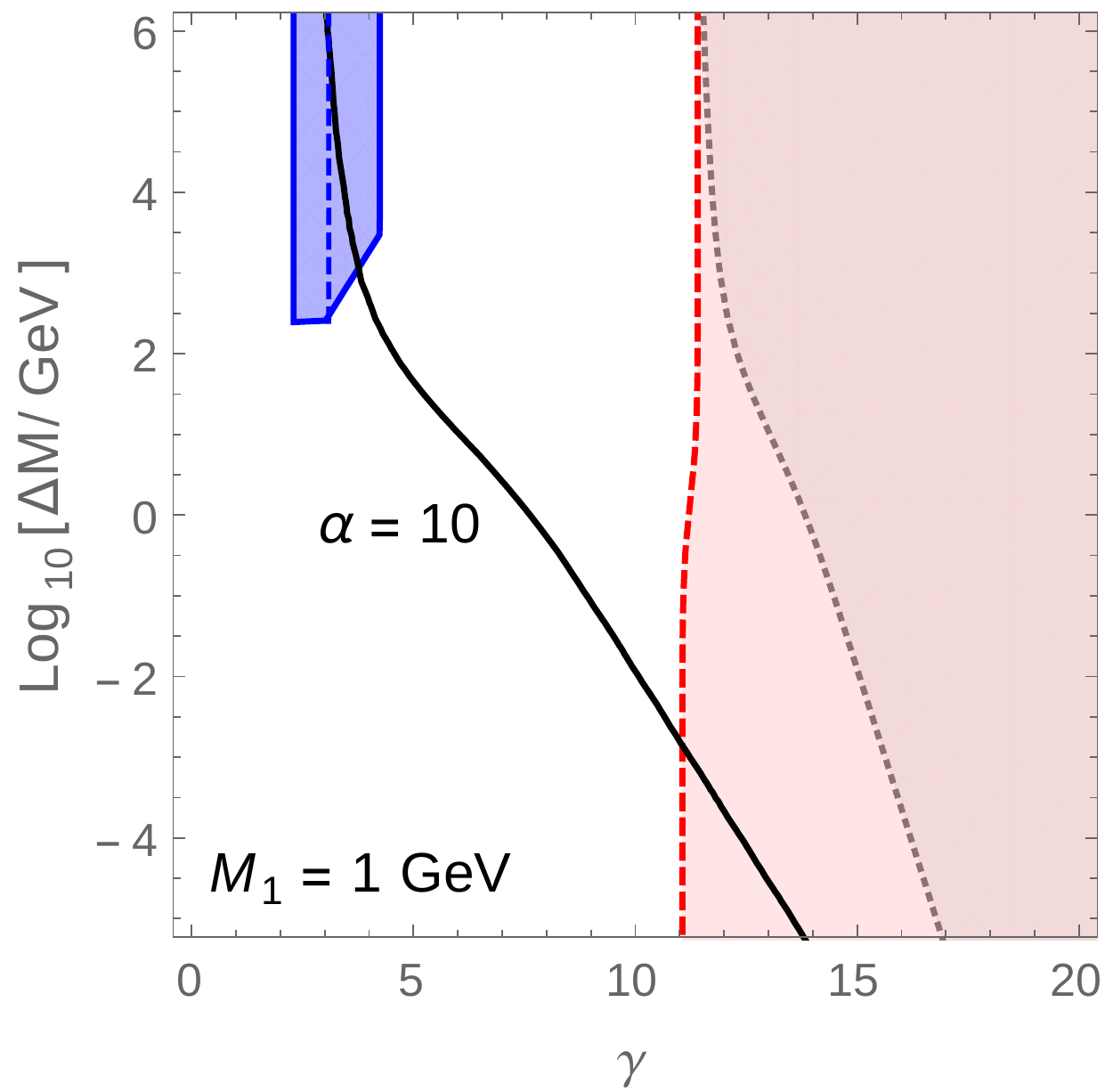}&
\includegraphics[width=0.33\textwidth,angle=0]{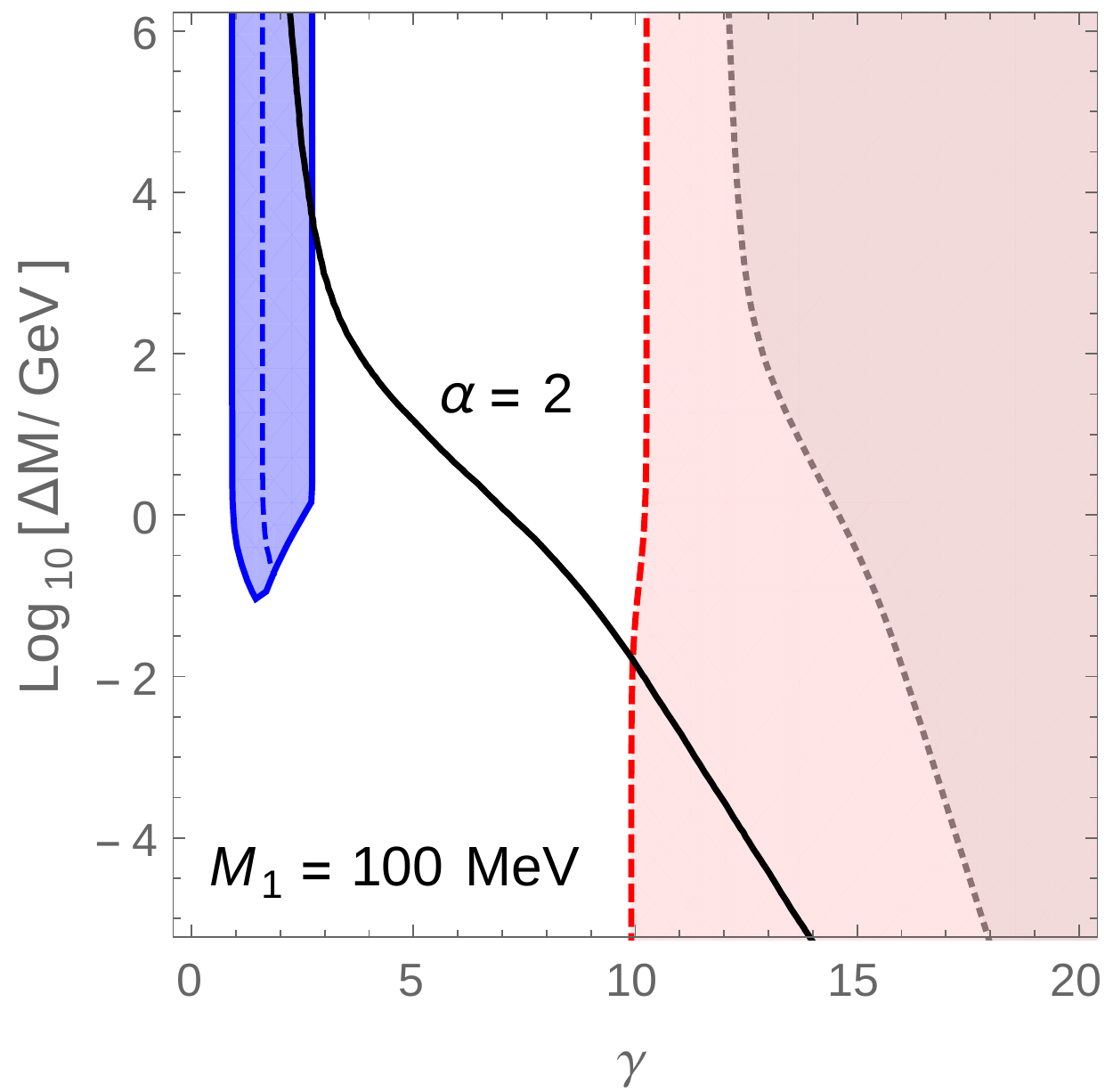}
 \end{tabular}
\caption{\label{fig3}{\small\mathversion{bold}
\textbf{Neutrinoless double beta decay ($M_1< 100~\text{GeV}$).} \mathversion{normal}
The same conventions as in Fig.~\ref{fig2}, but for different choices of  $M_1$.}}
\end{figure}

In the top panels (down panels) of Figs.~\ref{fig2} and \ref{fig3}, the blue shaded area corresponds to the region of the parameter space 
in which $10^{-2}~\text{eV}<|m_{\beta\beta}^{\rm light}\,+\,m_{\beta\beta}^{\rm heavy}|<0.5~\text{eV}$ ($10^{-2}~\text{eV}<|m_{\beta\beta}^{\rm heavy}|<0.5~\text{eV}$), 
projected on the $\gamma -\Delta M$ plane for NH (IH) and several values of $M_1$. In these plots we have fixed the already measured PMNS parameters
and neutrino oscillation mass differences to the best fit values given in \cite{nufit}. 
The relevant Majorana and Dirac CP violation phases 
in the PMNS matrix have been set to zero, but we have checked
that there is no significant impact on the results when other values 
are considered.
 The Casas-Ibarra parameter $\theta_{45}$ is also 
set to zero. It is irrelevant when the heavy Majorana neutrino 
exchange contribution is dominant (subdominant) in $m_{\beta\beta}$, 
but can play an important role in the interplay of the light and heavy 
Majorana neutrino exchange contributions when these two contributions are 
comparable in size \cite{Ibarra:2011xn}.
The Higgs mass has been fixed to $M_H=125$ GeV. The solid black line stands for different values, stated in the plots, of the $\alpha$ parameter defined as
\be
\alpha \equiv |m^{\rm 1-loop}_{\beta\beta}|/|m_{\beta\beta}^{\rm light}|\,,
\ee
where $m_{\beta\beta}^{\rm light}=m^{\rm tree}_{\beta\beta}+m^{\rm 1-loop}_{\beta\beta}$ is the full (tree-level plus one-loop)
contribution to $m_{\beta\beta}$ given by the light neutrinos.
Therefore, $\alpha$ quantifies 
the level of fine-tuning in the cancellation between $m^{\rm tree}_{\beta\beta}$ and $m^{\rm 1-loop}_{\beta\beta}$ described 
in section \ref{CIloop} and required in order to keep the light neutrino masses and mixing to the observational values. Notice that
the level of fine-tuning increases with $\alpha$. The region to the right of the black solid line corresponds to values of $\alpha$ larger  than those 
stated in the plots. 

In the red shaded area of Figs.~\ref{fig2} and \ref{fig3}, the ratio between the leading and next to leading order contributions to the light
neutrino masses in the seesaw expansion is smaller than $10$. The next to leading order contribution is given by~\cite{Grimus:2000vj}:
\be
\delta m_\nu = 
-\frac{1}{2}\left(m^{\rm tree}_\nu + m^{\rm loop}_\nu\right)\,
(\theta\,V) (\theta V)^\dagger\; -\; 
\frac{1}{2}\,(\theta V)^*\,(\theta V)^T\,
\left(m^{\rm tree}_\nu+m^{\rm loop}_\nu\right)\,.
\ee 
%
From this expression, one can conclude that a cancellation  
between the one-loop and tree-level contributions 
to the light neutrino masses remains
at next to leading order in the seesaw expansion. 
This is in agreement with Figs.~\ref{fig2}
and~\ref{fig3}, which show that the next to leading 
order contribution is always negligible in the range of parameters of interest. 

Ignoring for the time being
 the impact of the two-loop corrections, 
which will be commented below, two main conclusions can be extracted from 
Figs.~\ref{fig2} and \ref{fig3}. First, we have proved that a 
sizable and dominant heavy neutrino contribution to the 
$0\nu\beta\beta$ decay is possible for RH neutrino masses as 
heavy as $10$ TeV, satisfying at the same time the relevant 
constraints and keeping
under control the light neutrino mass and mixing pattern. 
Second, and not less important, it is shown that this possibility 
can only take place if a highly
fine-tuned cancellation between the tree-level and one-loop light 
neutrino masses is at work. The level of fine-tuning ranges 
from $\alpha=10^{4}$ to $10^{9}$,
for heavy masses between $M_1=100$ GeV and $M_1=10$ TeV. 
On the other hand, the level of fine-tuning is smaller 
for lighter masses, being in the case of $M_1=100$ MeV smaller than
$\alpha = 2$. In addition, we have checked that 
for $M_1\gtrsim 10$ TeV a heavy contribution to
$m_{\beta\beta}$  in the range of sensitivity
of the next-generation of experiments,
 $|m_{\beta\beta}|\gtrsim 0.01$ eV, 
cannot be expected.

Figs.~\ref{fig2} and \ref{fig3} also show that in the limit 
$\Delta M \gg M_1$ the sizable heavy neutrino contribution 
corresponding to the blue region becomes 
independent of $\Delta M$, according with the ESS 
limit -- see Eq.~(\ref{meeh2}). 
However, in the ISS limit $\Delta M \ll M_1$ 
this is not the case and, according to Eq.~(\ref{meeh2}), 
the smaller the heavy splitting $\Delta M$, the larger is the value of $\gamma$.

Notice that in the IH case we have plotted only 
$m_{\beta\beta}^{\rm heavy}$ because $|m_{\beta\beta}^{\rm light}|$ 
is already in the planned range of sensitivity 
of the next generation of $0\nu\beta\beta$ decay experiments.
 In this case for  
$M_1\lesssim 10^3$ GeV and $\Delta M << M_{1,2}$, there can be, in principle, 
a significant interplay between the light and heavy Majorana neutrino exchange contributions in 
the effective Majorana mass, as discussed at tree level in detail in \cite{Ibarra:2011xn} and summarised by us  
at the end of subsection II.A (see the paragraph before the last in subsection II.A). 
More specifically, due to this interplay of the light and heavy Majorana neutrino contributions,
$|m_{\beta\beta}|$ can be larger (smaller) than that predicted in the case of the exchange of light neutrinos
with IH mass spectrum and $|m_{\beta\beta}|$ will exhibit a dependence on the 
atomic number $A$ of the decaying nucleus. This can happen roughly in the region located to the left of the 
blue dashed line corresponding to $|m_{\beta\beta}^{\rm heavy}|=0.05$ eV inside the blue areas in Figs. \ref{fig2}
and \ref{fig3}.

In the NH case, the light neutrino contribution 
is smaller than $10^{-2}$ eV and 
therefore any sizable effect to the process is due to the heavy neutrinos. This is why in the NH case we plot 
the total contribution $m_{\beta\beta}$, including light and heavy neutrinos.
%

It follows from Fig. 4 that  for $M_1\geq 100$ GeV the regions of interest 
(the blue shaded areas) correspond to $\gamma \gtrsim 6$. 
For such values of $\gamma$, as it is not difficult to show, we have for the 
NH and IH neutrino mass spectra:
\bea
&M_1& |(\theta V)_{e1}|^2 \approx M_2 |(\theta V)_{e2}|^2 
\nonumber \\[0.30cm]
&\approx& \frac{e^{2\gamma}}{4}\,
\left |U_{e2}\sqrt{m_2} -\,i\, U_{e3}\sqrt{m_3} \right |^2\,,~~{\rm NH} 
\label{thVgammaNH1}
\\ [0.30cm]
&\approx& \frac{e^{2\gamma}}{4}\,
\left |U_{e1}\sqrt{m_1} -\,i\, U_{e2}\sqrt{m_2} \right |^2\,.~~{\rm IH} 
\label{thVgammaIH1}
\eea
%
Taking into account that Fig. 4 is obtained by setting to zero 
the phase $\theta_{45}$ and the Dirac and Majorana phases in the PMNS matrix
and by using the best fit values of the neutrino oscillation parameters, 
Eqs. (\ref{thVgammaNH1}) and  (\ref{thVgammaIH1}) imply the following relations 
between $|(\theta V)_{e1 (e2)}|^2$ and the parameter $\gamma$:
\be
M_1 |(\theta V)_{e1}|^2 \approx M_2 |(\theta V)_{e2}|^2 
\approx e^{2\gamma}\, 0.94~(12.4)\times 10^{-3}~{\rm eV}\,,~~{\rm NH~(IH)}\,. 
\label{thVgammaNHIH2}
\ee
%

In view of the high level of fine-tuning required in order to have a 
cancellation between the tree-level and one-loop light neutrino masses, 
the obvious question arising here is what is the role of 
the two-loop corrections. Can the two-loop corrections spoil 
this fine-tuned cancellation? In order to answer this question, 
we estimate the impact of the two-loop contributions. 
Since we are studying the case in which heavy neutrinos can 
give a sizable contribution to the $0\nu\beta\beta$ decay, which means 
relatively large
Yukawa couplings, we expect the diagram with two Higgs bosons 
in the loop to be the leading two-loop contribution to the 
light neutrino mass matrix. 
The contribution of this diagram can be roughly estimated as
\be
m^{\rm 2-loop}_{\beta\beta} \sim
\frac{y_{1e}^2}{(4\,\pi)^2}\,m^{\rm 1-loop}_{\beta\beta}\,,
\ee  
%
where $m^{\rm 1-loop}_{\beta\beta}$ is the one-loop contribution 
in $m_{\beta\beta}^{\rm light}$. This estimate of the impact of the two 
loop corrections is also shown in Figs.~\ref{fig2} and \ref{fig3}, where 
the gray area to the right of the dotted line corresponds 
to the region of the parameter space with 
$y_{1e}^2\, m^{\rm 1-loop}_{\beta\beta}> 16\,\pi^2\,m_{\beta\beta}^{\rm light}$. 
This region of the parameter space is excluded since the 
two-loop correction, which would dominate the light neutrino masses, 
would be larger than the value dictated by neutrino oscillation data. 
Notice that this would essentially exclude the possibility of having a 
large sterile neutrino contribution for $M_1 \gtrsim 1$ TeV, 
as can be seen in Figs.~\ref{fig2} and \ref{fig3}. 
For $M_1 \lesssim 100$ GeV  the impact of the two-loop correction is 
basically negligible.

\section{Conclusions}

We have performed a systematic analysis of the radiative corrections 
to the light  neutrino masses arising in low 
scale type I seesaw scenarios, where the RH (sterile) neutrino masses 
vary in the interval $100~\text{MeV}\lesssim M \lesssim 10~\text{TeV}$.
Within this range of masses a significant enhancement of the 
neutrinoless double beta ($0\nu\beta\beta$)  decay rate 
in several isotopes - at the level of
sensitivity of the present and next generation experiments searching for
this rare process - is possible, due to the new physics contribution 
in the decay amplitude given by the exchange of the 
virtual heavy sterile neutrinos. Notice that one of the most clear signatures of a significant 
heavy sterile Majorana neutrino contribution to the 
$0\nu\beta\beta$ decay amplitude is the dependence of the 
$0\nu\beta\beta$ decay effective Majorana mass,
$|m_{\beta\beta}|$, on the atomic number $A$ of the 
decaying nucleus \cite{HPR83}.

The requirement of a sizable contribution of heavy neutrinos 
with masses $\gtrsim 1$ GeV 
to the $0\nu\beta\beta$ decay implies strong cancellations 
between the tree-level and one-loop expressions in the
light neutrino mass matrix $m_\nu$ originated from the seesaw mechanism. 
We show that such a cancellation can always be achieved while 
being consistent with neutrino
oscillation data and low energy constraints from direct searches, 
charged lepton flavour violation and non-unitarity by 
using a generalisation of the Casas-Ibarra parametrization of 
the neutrino Yukawa matrix, which can be derived from 
Eqs.~(\ref{CIloop1}) and (\ref{thetaR2}).
We clarify the connection between this parametrization and 
the lepton number breaking terms in the seesaw Lagrangian, 
as usually defined in extended as well as  inverse/direct 
seesaw UV completions of 
the Standard Model. Then, we numerically quantify the level of 
fine-tuning between 
the tree-level and one-loop parts of $m_\nu$ in the case the 
heavy neutrino contribution $m_{\beta\beta}^{\text{heavy}}$
to the effective Majorana neutrino mass - which enters in the $0\nu\beta\beta$ decay amplitude - 
is sizable, namely $|m_{\beta\beta}^{\text{heavy}}|\gtrsim 0.01$ eV. 

The main results of our analysis
are summarised in Figs.~\ref{fig2} and \ref{fig3}, 
where we show that a fine-tuning of one part in $10^{4}$ ($10^{5}$) 
for RH neutrino masses  $\sim 100$ ($1000$) GeV is unavoidable 
in order to have an observable effect in $0\nu\beta\beta$ decay experiments.
Furthermore, we conclude that for seesaw scales $M$ larger than few TeV, 
two-loop effects in the generation of the light neutrino masses cannot 
be neglected, thus 
excluding the possibility of having a large  $|m_{\beta\beta}^{\text{heavy}}|$.
Conversely, in the low mass regime,  $M\lesssim 1$ GeV,  the level 
of fine-tuning in the seesaw parameter space is  very mild and 
the sterile neutrino contribution can easily
exceed the current limits on the effective Majorana neutrino mass.

Finally, we can conclude on the basis of the results obtained 
in the present analysis that $0\nu\beta\beta$ sets the strongest constraints on lepton number 
violation in low scale type I seesaw extensions of the Standard Model. 
In particular, this implies a strong suppression of processes 
which involve the production  at colliders (LHC included) 
of RH neutrinos and their decays with two like-sign charged leptons 
in the final state (see, e.g., \cite{delAguila:2008cj, Ibarra:2010xw}).

\acknowledgments
The work of J.L.P. and S.T.P. was supported in part by the European Union FP7 
ITN INVISIBLES (Marie Curie Actions, PITN-GA-2011-289442-INVISIBLES),
by the INFN program on Theoretical Astroparticle Physics (TASP) 
and by the research grant  2012CPPYP7 
({\it  Theoretical Astroparticle Physics})
under the program  PRIN 2012 funded by the Italian 
Ministry of Education, University and Research (MIUR). 
S.T.P. acknowledges partial support 
from the World Premier International Research Center
Initiative (WPI Initiative), MEXT, Japan.

\appendix

\mathversion{bold}
\section{$m_{\nu}^{\rm 1-loop}$ in an arbitrary basis for 2 RH neutrinos}
\mathversion{normal}\label{App}

We report in this appendix the full computation of the one-loop correction (\ref{mv1loopB}) to the light neutrino mass matrix 
 in terms of the seesaw parameters introduced in Eq.~(\ref{Mnu2}),
from which it is possible to derive the one-loop correction to effective Majorana neutrino mass in the extended and inverse seesaw limits, Eqs.~(\ref{meff1loopESS}) and 
(\ref{meff1loopISS}), respectively.
In order to obtain an analytic expression for the one-loop neutrino mass matrix, we conveniently change the basis of the heavy RH neutrinos, i.e.
$\nu_{aR}= \hat{V}_{ab} \,\nu_{bR}^{\prime}$, with the unitary transformation
\begin{eqnarray}
\hat{V} & =& \frac{1}{\sqrt{2}} \left(
     \begin{array}{cc} 
   i & 1  \\
   -i & 1
    \end{array}\right)\,.\label{basis2}
\end{eqnarray}
In the new basis the RH neutrino Majorana mass matrix takes the form:
\begin{eqnarray}
M_{R}^{\prime} &\equiv & \hat{V}^{T}\,M_{R}\,\hat{V} \,=\; \frac{1}{2}\left(
     \begin{array}{cc} 
  2 \,\Lambda -(\mu+\mu^{\prime}) & -i (\mu-\mu^{\prime})  \\
   -i (\mu-\mu^{\prime}) & 2\, \Lambda + (\mu+\mu^{\prime})\,.
    \end{array}\right)
\end{eqnarray}
Then, the resulting one-loop Majorana mass term for active neutrinos is 
 \begin{equation}
 	m_{\nu}^{\rm 1-loop}\;=\; 
	\frac{1}{(4\pi v)^{2}}\,m_{D}\,\hat{V}\,\left({M_{R}^{\prime}}^{-1}\,F(M_{R}^{\prime}{M_{R}^{\prime}}^{\dagger})+F({M_{R}^{\prime}}^{\dagger}M_{R}^{\prime})\,{M_{R}^{\prime}}^{-1} \right) \,\hat{V}^{T}\,m_{D}^{ T}\,,\label{mv1loopC}
 \end{equation}
where the loop function $F(x)$ is defined in Eq.~(\ref{1loopfunc}) and the Dirac mass matrix $m_D$ is parametrized as in (\ref{Mnu2}).
In this case we have:~\footnote{We assume without loss of generality that the parameter  $\Lambda$ in (\ref{Mnu2}) is real.}
\begin{eqnarray}
	&&{M_{R}^{\prime}}\,{M_{R}^{\prime}}^{\,\dagger}\;=\;
	\left(\Lambda^{2}+\frac{1}{2}\left(|\mu|^{2}+|\mu^{\prime}|^{2}\right)\right)\left(\mathbf{1_{2\times 2}}
	-A(a,b,c)\right)\,,\\
	&&{M_{R}^{\prime}}^{\,\dagger}\,M_{R}^{\prime}\;=\;
	\left(\Lambda^{2}+\frac{1}{2}\left(|\mu|^{2}+|\mu^{\prime}|^{2}\right)\right)\left(\mathbf{1_{2\times 2}}
	-A(a,-b,c)\right)\,,
\end{eqnarray}
where
\begin{equation}
A(a,b,c)\equiv\left(a\,\mathbf{\sigma_{3}}\,+\,b\,\mathbf{\sigma_{2}}\,+\,c\,\mathbf{\sigma_{1}}\right)\,,  
\end{equation}
$\mathbf{\sigma}_{i}$ ($i=1,2,3$) denoting the $2\times 2$ Pauli matrices. 
The real parameters $a$, $b$ and $c$ are defined~as
\begin{eqnarray}
	a &=& \frac{2\, \Lambda\,{\rm Re}\left(\mu+\mu^{\prime}\right)}{2\,\Lambda^{2}\,+\,|\mu|^{2}\,+\,|\mu^{\prime}|^{2}}\,,\\
	b &=& \frac{|\mu^{\prime}|^{2}-|\mu|^{2}}{2\,\Lambda^{2}\,+\,|\mu|^{2}\,+\,|\mu^{\prime}|^{2}}\,,\\
	c &=& \frac{2\,\Lambda\,{\rm Im}\left(\mu^{\prime}-\mu\right)}{2\,\Lambda^{2}\,+\,|\mu|^{2}\,+\,|\mu^{\prime}|^{2}}\,.
\end{eqnarray}
In this way, one can obtain a closed form for the logarithms which enter in Eq.~(\ref{mv1loopC}) through the loop function.
Indeed, we have
\begin{eqnarray}
	\log\left[\mathbf{1_{2\times 2}}-A(a,b,c)\right] &=&-\, \sum\limits_{n=1}^{\infty}\frac{1}{n} 
	A(a,b,c)^{n}\,,\label{sum1}
\end{eqnarray}
with 
\begin{eqnarray}
	\left[A(a,b,c)^{n}\right]_{11} &=&  \frac{1}{2}  \left(a^{2}+b^{2}+c^{2}\right)^{\frac{1}{2}(-1+n)}\left( a\,(1-(-1)^{n})+\sqrt{a^{2}+b^{2}+c^{2}}\,(1+(-1)^{n})\right)\,,\nonumber\\
	\left[A(a,b,c)^{n}\right]_{22} &=&  -\frac{1}{2}  \left(a^{2}+b^{2}+c^{2}\right)^{\frac{1}{2}(-1+n)}\left( a\,(1-(-1)^{n})-\sqrt{a^{2}+b^{2}+c^{2}}\,(1+(-1)^{n})\right)\,,\nonumber\\
	\left[A(a,b,c)^{n}\right]_{12} &=& \left[A(a,b,c)^{n}\right]_{21}^{*} \,=\, \frac{i}{2}\left(b+i\,c\right) \left(a^{2}+b^{2}+c^{2}\right)^{\frac{1}{2}(-1+n)}\left(-1+(-1)^{n}\right)\,.
\end{eqnarray}
Then, one can show that the infinite series in (\ref{sum1}) gives the exact results
\begin{eqnarray}
	\big[\log\left[\mathbf{1_{2\times 2}}-A(a,b,c)\right]\big]_{11} &=&  \frac{2 \,a \tanh^{-1}\left(\sqrt{a^{2}+b^{2}+c^{2}}\right)}{\sqrt{a^{2}+b^{2}+c^{2}}}-\log\left(1-a^{2}-b^{2}-c^{2}\right)\,, \nonumber\\
	\big[\log\left[\mathbf{1_{2\times 2}}-A(a,b,c)\right]\big]_{12} &=& \left[\log\left(\mathbf{1_{2\times 2}}-A(a,b,c)\right)\right]_{21}^* = \frac{i\left(b+i\,c\right)\log\left(\frac{1-\sqrt{a^{2}+b^{2}+c^{2}}}{1+					\sqrt{a^{2}+b^{2}+c^{2}}}\right)}{\sqrt{a^{2}+b^{2}+c^{2}}}\,, \nonumber\\
	\big[\log\left[\mathbf{1_{2\times 2}}-A(a,b,c)\right]\big]_{22} &=& -\frac{2\, a \tanh^{-1}\left(\sqrt{a^{2}+b^{2}+c^{2}}\right)}{\sqrt{a^{2}+b^{2}+c^{2}}}-\log\left(1-a^{2}-b^{2}-c^{2}\right)\,.\;\;\;\;\;\;\;\;\label{A12}
\end{eqnarray}
Therefore, by replacing Eqs.~(\ref{A12}) in  (\ref{mv1loopC}), we obtain an analytic expression for the one-loop contribution to the light Majorana neutrino mass matrix 
as a function of the parameters given in (\ref{Mnu2}).

  \end{document}